\documentclass[aps,nofootinbib,nolongbibliography,superscriptaddress,twocolumn,prd,10pt]{revtex4-2}
\usepackage[T1]{fontenc} 
\usepackage{amsmath,amssymb}
\usepackage{graphicx}
\usepackage{dcolumn}
\usepackage{bm}
\usepackage[normalem]{ulem}
\usepackage{hyperref}

\usepackage{xcolor}
\usepackage{natbib}
\usepackage{xspace}

\definecolor{Rayne}{HTML}{0264a1}

\definecolor{DG}{HTML}{FF0000}

\newcommand{\ax}{a}
\newcommand{\iso}{{\rm iso}}
\newcommand{\osc}{{\rm osc}}

\newcommand{\ini}{{\rm ini}}

\newcommand{\Hinf}{H_{\rm inf}}

\newcommand{\dm}{{\rm dm}}
\newcommand{\de}{{\rm de}}
\newcommand{\eq}{{\rm eq}}
\newcommand{\axionCAMB}{\textsc{axionCAMB}\xspace}
\newcommand{\axie}{\textsc{AxiECAMB}\xspace}
\newcommand{\camb}{\textsc{CAMB}\xspace}
\newcommand{\ULA}{axion\xspace}
\newcommand{\Mpl}{M_{\rm pl}}

\begin{document}

\title{Inflationary Axion Isocurvature in the CMB across All Ultralight Masses
}

\author{Rayne Liu}\email{rayneliu@uchicago.edu}
\affiliation{Kavli Institute for Cosmological Physics, Enrico Fermi Institute, and Department of Astronomy \& Astrophysics, University of Chicago, Chicago, IL 60637
}

\author{Wayne Hu}
\affiliation{Kavli Institute for Cosmological Physics, Enrico Fermi Institute, and Department of Astronomy \& Astrophysics, University of Chicago, Chicago, IL 60637
}

\author{Daniel Grin}
\affiliation{Department of Physics and Astronomy, Haverford College, 370 Lancaster Avenue, Haverford, PA 19041
}

\date{\today}
\begin{abstract}
If the Peccei-Quinn symmetry of an ultralight axion is broken before the end of inflation, 
axion quantum fluctuations seed isocurvature perturbations, linking them to the tensor-to-scalar ratio $r$.
 We extend the effective time average (ETA) approach of the Boltzmann code \axie to accurately evolve these perturbations across the full axion mass range from dark energy ($m_\ax \lesssim H_0$) to dark matter ($m_\ax \gg 10^{-28}$ eV) types. We provide analytic fitting formulae for the axion abundance given the initial field value $\phi_\ini$, accurate at sub-percent level for $m_\ax \gg H_0$ and allowed dark matter fraction $f_\dm$.  In the dark matter regime, the Planck bound on CDM isocurvature requires $r f_\dm < 0.08\,(m_\ax/10^{-27}\,{\rm eV})^{-1/2}$, which becomes stronger than the current BICEP tensor bound for $m_\ax f_\dm^2 \gtrsim 10^{-26.4}\,{\rm eV}$.  For $10^{-32} \lesssim m_\ax/{\rm eV} \lesssim 10^{-28}$, Jeans suppression breaks the degeneracy with CDM isocurvature, leaving unique signatures, 
and in the dark energy regime ($m_\ax \lesssim H_0$), the isocurvature signal is even more highly suppressed, peaking only at the CMB quadrupole. We provide analytic scalings for both signatures.
Given the tensor bound, any primary CMB detection in these two lightest regimes would indicate a non-inflationary origin of the isocurvature modes or a breakdown of the standard frozen-field misalignment scenario. A window of coexistence opens near $m_\ax \sim 10^{-25}$\,eV and $f_\dm \gtrsim 0.1$ where both axion isocurvature and tensor modes could be discovered just below current bounds while simultaneously alleviating the $S_8$ tension.
\end{abstract}

\maketitle

\section{\label{sec:intro}Introduction}

Scalar and pseudoscalar fields are ubiquitous in  beyond-the-standard-model physics.
For example, in string theory, the dilaton is a scalar field, while axion-like pseudoscalars can emerge as Kaluza-Klein zero modes of antisymmetric tensor fields compactified on Calabi-Yau manifolds and can couple to the electroweak sector or to new gauge sectors \cite{Svrcek:2006yi}.
One of the pseudoscalars might be the quantum chromodynamics (QCD) axion, while many more could populate a wide range of axion masses, approximately uniformly in $\log m_\ax$. In this work, we focus on the ultralight mass range of such fields, including scalar degrees of freedom associated with vector fields, and refer to them generically as ``axions'' for simplicity.

Regardless of their fundamental origin, if the axion exists and is light during inflation, $m_\ax \ll \Hinf$, where $\Hinf$ denotes the Hubble rate, it will acquire quantum field fluctuations of order $\Hinf/2\pi$ and generate an isocurvature component in cosmic initial conditions \cite{Axenides:1983hj,Turner:1983sj,Seckel:1985tj,Lyth:1991ub,Fox:2004kb,Hertzberg:2008wr,WMAP:2008lyn,Marsh:2013taa,Marsh:2014qoa,Visinelli:2014twa,Hlozek:2017zzf}.  For the Peccei-Quinn (PQ) axion this would require  PQ symmetry breaking to have occurred before the end of inflation with a decay constant $f_\ax > \Hinf$.  If we further assume the axion field acquires a value $\phi_\ini \ll f_\ax$, then its subsequent evolution is in an effectively quadratic potential.  
Once the Hubble rate drops to $H \sim m_\ax$ after inflation, the axion oscillates coherently in this potential with angular frequency $\omega\sim m_\ax$  and its density redshifts as cold dark matter.  In this regime the axion can comprise part of the dark matter. 
If $m_\ax/H_0<10$, these coherent oscillations have begun only recently, if at all, and the axion can compose part or all of the dark energy.

Across this wide range of masses, axions produce many distinct signatures.  Axions could also be a fractional component of the dark matter or dark energy in the universe: CMB, LSS, and Lyman-$\alpha$ observations impose constraints on the \ULA fractional abundance of $f_\dm=\Omega_{a}/\Omega_{\dm}\lesssim 0.05$ when $10^{-32}~{\rm eV}\lesssim m_\ax\lesssim 10^{-25}~{\rm eV}$ \cite{2011MNRAS.415L..40B,Hlozek:2014lca,2016A&A...591A..58P,Hui:2016ltb,Hlozek:2017zzf,Irsic:2017yje,2019ApJ...871...28B,Rogers:2020ltq,Dalal:2020mjw,Dentler:2021zij,Lague:2021frh,Rogers:2023ezo,Winch:2024mrt,Sipple:2024svt}.
In this work we develop accurate and precise methods to calculate and test the cosmological signatures of these isocurvature perturbations in the linear or large-scale regime, improving  upon past results for CMB signatures, constraints, and forecasts for future CMB experiments 
\cite{Axenides:1983hj,Seckel:1985tj,Visinelli:2014twa,Marsh:2014qoa,Marsh:2013taa,Fox:2004kb,WMAP:2008lyn,Planck:2018jri,Gordon:2004ez,Coble:1996te,Frieman:1995pm,Hertzberg:2008wr,Lyth:1991ub,Hlozek:2017zzf,Batell:2026avi}.
These considerations make the mass range $m_\ax \lesssim 10^{-24}$\,eV of particular interest here, though our techniques also provide initial conditions for nonlinear simulations of structure for even heavier masses within the fuzzy dark matter regime.

Most prior work testing axion masses in this regime used the Boltzmann code \axionCAMB.\footnote{\url{https://github.com/dgrin1/axioncamb}} This package is a modification to the \camb Boltzmann code,\footnote{\url{https://camb.info/}} which implemented perturbation evolution with an effective-fluid approximation (EFA) where a sharp switch is made from field to fluid equations when $m_\ax> 3H$.
This approximation suffers from an artificial dependence on the phase of the field oscillation and other artifacts due to the implementation of a hard switch.  These artifacts can impact observables and thus constraints \cite{Cookmeyer:2019rna,Passaglia:2022bcr}.

Improved methods were introduced in Ref.~\cite{Passaglia:2022bcr} that involved matching the EFA to the effective time average (ETA) of the field oscillations  (see \cite{Urena-Lopez:2015gur} for an alternative approach).
 These methods were extended into a full Boltzmann treatment for adiabatic perturbations in Ref.~\cite{Liu:2024yne}, and the resulting code, \axie, replaces 
 \axionCAMB as the reference \camb implementation\footnote{\url{https://github.com/Ra-yne/AxiECAMB}} for axions (see also \cite{Baryakhtar:2024rky,Moss:2025ymr,2026arXiv260512054G}  for other implementations of the ETA approach and \cite{Liu:2025bss,2026arXiv260606410L} for emulators of existing codes).  Here we further extend these methods to axion isocurvature fluctuations and address the additional requirements for their accurate calculation due to their strong sensitivity to the dynamics of the axion system.

 Constraints on axion isocurvature fluctuations can then be translated into constraints on $\Hinf$ or equivalently the tensor-to-scalar ratio $r$.   In order to complete this mapping, we need an accurate calibration of the initial field value $\phi_\ini$ corresponding to a given present-day axion density or abundance.  Since this requires an accurate transition between the field and fluid regime, our ETA technique is especially useful here and we provide a simple expression that is accurate across the full range of masses.   Our abundance expression is useful for relic axion studies even beyond the isocurvature context here.

 In summary, we: (i) extend \axie to inflationary axion isocurvature modes; (ii) derive and calibrate analytic fits connecting the initial field value $\phi_\ini$ to the axion abundance $\Omega_\ax h^2$, and the isocurvature amplitude to the tensor-to-scalar ratio $r$ over the full mass range; (iii) quantify CMB signatures in three mass regimes (dark-matter-like, intermediate, and dark-energy-like axions) independently of their inflationary origin; and (iv) translate isocurvature constraints into bounds on the tensor-to-scalar ratio $r$ through the inflationary Hubble scale $\Hinf$, highlighting the parameter region where isocurvature and tensors can coexist near current bounds while also addressing implications for the $S_8$ tension.

The outline of this paper is as follows.  We begin in Sec.~\ref{sec:ic} with a review of axion isocurvature fluctuations and present the analytic relationships between inflationary and late-time parameters as calibrated by  \axie.   In Sec.~\ref{sec:cmb}, we calculate and describe the isocurvature signatures in the CMB and matter power spectra in  the three distinct mass regimes deriving bounds on the isocurvature amplitude.  In Sec.~\ref{sec:discussion}, we discuss these results.   In the Appendix we detail the code considerations for the new release of \axie\,v1.1, including accuracy tests.

For illustrating the observables, we take a fiducial spatially flat $\Lambda$CDM cosmology  with \emph{Planck} 2018 baseline parameters: baryon density $\Omega_{b}h^2=0.0224$, sum of the neutrino masses $\sum m_\nu = 0.06\,$eV, Hubble constant
 $H_0 = 67.36$\,km/s/Mpc, reionization optical depth $\tau=0.0544$, amplitude of the curvature spectrum $\ln(10^{10}A_s) = 3.044$ with a tilt $n_s=0.9649$  \cite{Aghanim:2018eyx}. In the presence of axions with $m_\ax/H_0\geq 10$, we replace the cold dark matter density $\Omega_c h^2$ with the total dark matter $\Omega_\dm h^2=\Omega_c h^2+\Omega_\ax h^2=0.12$, where $\Omega_\ax h^2$ is effectively time-averaged over the mass oscillation timescale. For $m_\ax< 10 H_0 =1.44\times 10^{-32}$\,eV axions behave as dark energy (DE). Unless otherwise specified, we parametrize  its abundance  with $f_\de\equiv \Omega_\ax/\Omega_\de $, where
 $\Omega_\de  = \Omega_\Lambda+\Omega_\ax $, where $\Omega_\ax$ is the instantaneous value today, and keep $\Omega_c h^2=0.12$.  For illustration purposes we take a scale-invariant isocurvature spectrum $n_\ax=0$, consistent with tensor upper bounds, though our treatment is general and applies to any spectrum.
     Throughout this work we  utilize units where $c = \hbar = 1$,  overdots denote derivatives with respect to the conformal time, and adopt synchronous gauge as in the \camb family of codes.

\section{Axions from Inflation} \label{sec:ic}
If Peccei-Quinn symmetry is broken prior to the end of inflation, axions carry quantum vacuum fluctuations with an rms amplitude $\Hinf/2\pi$ per e-fold, seeding a field power spectrum 
\begin{align}
\Delta^2_{\delta \phi_\ini} \equiv{} & \frac{k^3 P_{\delta \phi_\ini}}{2\pi^2} =\left. \left( \frac{\Hinf}{2\pi} \right)^2 \right|_{\frac{k}{a H}|_{\rm inf}=1}\nonumber\\
={}& \frac{A_s r_{\ax}}{8} \left(\frac{k}{k_0}\right)^{n_\ax} \Mpl^2.\label{eq:fielddimspec}
\end{align}
Fluctuations on scales that exit prior to today's horizon scale contribute instead to the average initial field $\phi_\ini$ within our horizon.

Here we follow the standard tensor-power-spectrum convention for the tensor-to-scalar ratio $r$ and tensor tilt $n_T$. Since gravitational wave fluctuations are similarly generated by $\Hinf$, we define the isocurvature amplitude $r_{\ax}$ relative to the scalar amplitude $A_s$ at $k=0.05$\,Mpc$^{-1}$ and the tilt $n_\ax$ around that scale.  Although the tensor and isocurvature parameters must be equal in a consistent single-field inflationary model, we allow the freedom to vary $r_{\ax}$ and $n_{\ax}$ independently to identify their respective contributions to CMB fluctuations.   Since the tensor $r$ is already known to be small and $n_T=-r/8$ in canonical slow-roll inflation, when illustrating our results with numerical calculations we always take $n_\ax=0$. 

Since axions are a spectator field in the early universe, their field fluctuations represent isocurvature modes. 
Together with the fluctuation amplitude $r_\ax$, 
the background field value from inflation $\phi_\ini$, the field fluctuations set the initial isocurvature density power spectrum.  For an initial field that is sufficiently close to the potential minimum so that the Klein-Gordon (KG) equation,
\begin{equation}
\Box\phi= -m_\ax^2 \phi,
\end{equation}
applies, the background field is frozen by Hubble drag when $m_\ax\ll H$ and begins oscillating around its potential minimum around $m_\ax/H(a_\osc)=3$.
Consequently, the axion density abundance today obeys
\begin{equation}
\Omega_\ax H_0^2 = F_w m_\ax^2 \left(\frac{\phi_\ini}{\Mpl}\right)^2 a_\osc^3,
\end{equation}
where the exact proportionality coefficient $F_w$ can be obtained analytically for a background expansion history with a constant total equation of state $w=p/\rho$ at $a_\osc$.
For $m\gg H_\eq$, the background is radiation dominated (RD), $w=1/3$, and 
\begin{equation}
F_{1/3}= {}
\frac{4 (\Gamma[5/4])^2}{9\sqrt{3}\pi}, \quad
a_\osc = a_\eq \left( \frac{9}{2} \frac{H_\eq^2}{m_\ax^2} \right)^{1/4},
\label{eq:abundanceRD}
\end{equation}
where $a_\eq$ is the scale factor at matter-radiation equality and \begin{align}
H_\eq ={}& \sqrt{2}H_0(\Omega_\dm+\Omega_b)^{1/2} a_\eq^{-3/2}\nonumber\\
\approx{}& 2.19 \times 10^{-28}\left(\frac{\Omega_\dm h^2+\Omega_b h^2}{0.14}\right)^2 {\rm eV} 
\end{align}
is the corresponding Hubble rate, counting massive neutrinos as radiation at $a_\eq$ and matter at the present.
Combining these expressions, $\Omega_\ax h^2 \propto \phi_\ini^2 m_\ax^{1/2}$ in this radiation dominated regime.\footnote{The commonly used approximations that the field is frozen (e.g.~\cite{Marsh:2014qoa}) or obeys the field equation  until $a_\osc$ and thereafter redshifts like matter (e.g.~\cite{Hlozek:2014lca})  correctly give this scaling but differ in abundance normalization by ${\cal O}(1)$ \cite{Passaglia:2022bcr} (see also Appendix \ref{app:accuracy}).
}

For $H_0 \ll m_\ax \ll H_\eq$ and $\Omega_\ax \ll \Omega_c+\Omega_b+\Omega_\nu \approx \Omega_m$, the background is matter dominated (MD), $w=0$ and
\begin{align}
F_0  &=\frac{1}{24},\quad 
a_\osc =\left( \frac{9\Omega_m H_0^2}{m_\ax^2}\right)^{1/3}.
\label{eq:F0}
\end{align}
Therefore $\Omega_\ax h^2 \propto \phi_\ini^2$ with no $m_\ax$ dependence.
Only for axions in this mass regime do our analytic solutions  require $\Omega_\ax\ll \Omega_m$.  This condition is also imposed by current observational constraints. In Appendix \ref{app:accuracy}, we show that for the allowed $\Omega_\ax \lesssim 0.05\Omega_m$ this scaling works at the percent level or better.

For $m_\ax < H_0$, or dark energy (DE) domination, the field does not roll before $a=1$ and
\begin{align}
     \Omega_\ax H_0^2  = \frac{1}{6}  m_\ax^2 \left(\frac{\phi_\ini}{\Mpl}\right)^2.
     \label{eq:abundanceDED}
  \end{align}
Here the abundance scales steeply with mass $\Omega_\ax h^2 \propto \phi_\ini^2 m_\ax^2$.

We use the numerical solutions from \axie to interpolate between these scaling behaviors across the full mass range.  \axie implements the highly accurate ETA approach of Ref.~\cite{Passaglia:2022bcr} where the KG background solutions are matched to a time-averaged fluid description at a given switch time when the Hubble rate drops to $H_*$.
Fractional errors in the abundance are substantially smaller than $(H_*/m_\ax)^2$.    \axie v1.1 now provides $\phi_\ini$ for any given abundance and mass.  

When adiabatic modes are present, as they must be in our Universe, the default value of the switch, $m_\ax/H_*=10$, is sufficient given allowed amplitudes for isocurvature modes and cosmic variance. In this work,
to  evaluate the  axion isocurvature modes to high accuracy relative to their already  small contribution, we raise the switch to $m_\ax/H_* = 20$ and the accuracy settings
(see Appendix~\ref{app:accuracy} for further details).

The three asymptotic limits possess simple power-law scalings, but to extend the validity across the full mass range,
we can accurately interpolate between them and match the numerical solution from \axie:    
\begin{equation}\label{eq:F}
\Omega_\ax H_0^2 =F\left[\lg \left(\frac{m_\ax}{H_\eq}\right)\right] m_\ax ^2 \left(\frac{\phi_\ini}{\Mpl}\right)^2 a_\osc^{3},
\end{equation}
where $\lg$ denotes $\log_{10}$,
\begin{equation}\label{eq:Ffit}
F(x) = F_0 + \frac{F_{1/3}-F_0}{2} \left[ 1+ \tanh\left(\frac{x-x_0}{d}\right)\right],
\end{equation}
with $x_0=0.23$, $d=1.52$ 
and $a_\osc$ is defined piecewise as the solution to
\begin{align}\label{eq:aosc}
\frac{\Omega_m}{a_\osc^{3}} + \frac{(\Omega_\dm+\Omega_b){a_\eq}}{{a_\osc^{4}}}+C={}& \frac{m_\ax^2}{9H_0^2}, &&\!\!\!\!\!\  m_\ax \ge m_{\de},\nonumber\\
a_\osc^3 ={}& 4, &&\!\!\!\!\!\  m_\ax < m_{\de}.
\end{align}
Here $m_{\de}$ is the mass for which $a_\osc^3=4$, and in the fiducial model $m_\de \approx 1.58H_0$. The value 4 is chosen so that all lighter masses return the frozen-field abundance of Eq.~(\ref{eq:abundanceDED}) with $F\approx F_0=1/24$. The constant $C$ is adjusted so as to join the  oscillating and frozen regimes at the present epoch. For $\Lambda$CDM cosmologies that fit current data, $C=0.2$ suffices.  

\begin{figure}
    \centering
    \includegraphics[width=1\linewidth]{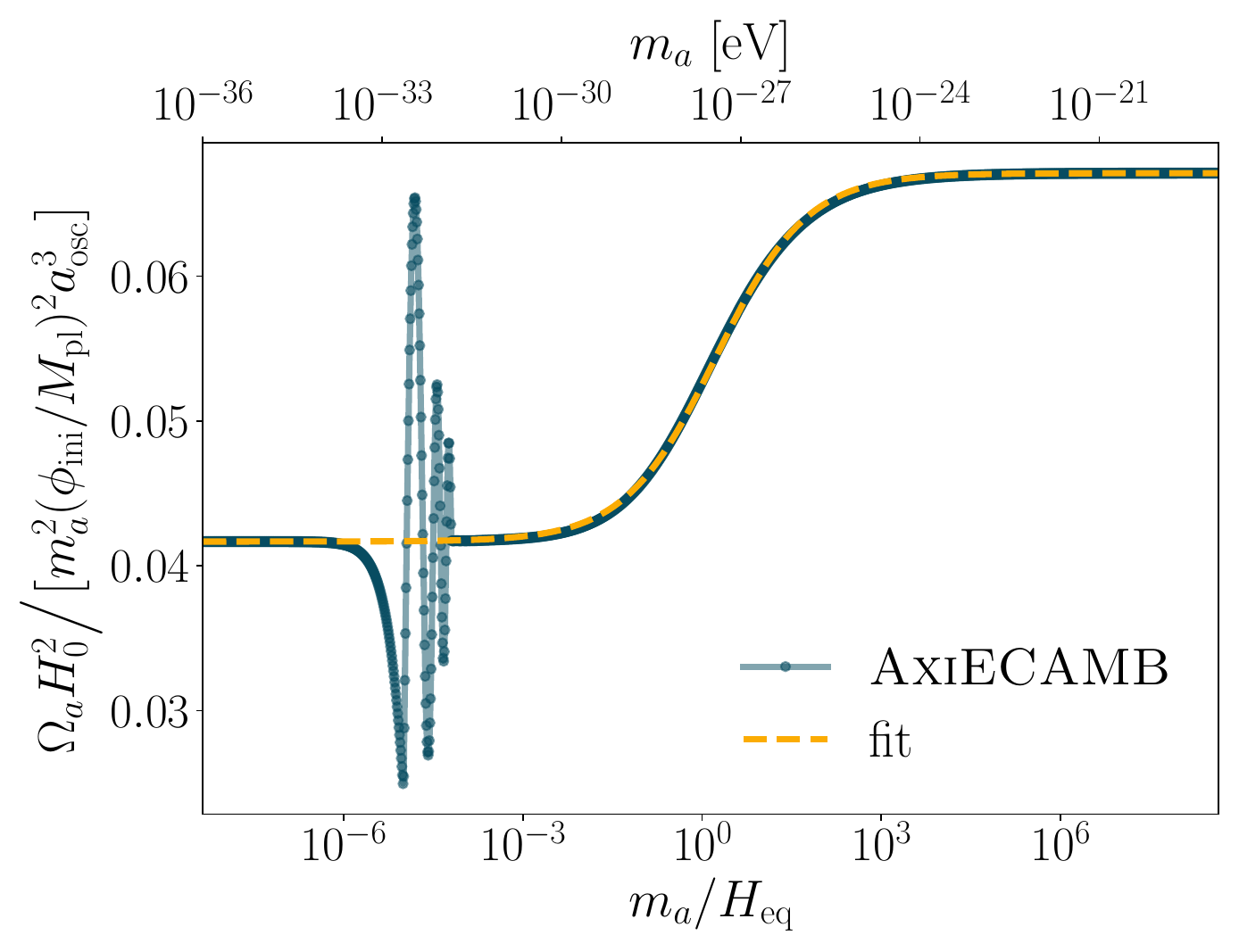}
    \caption{Axion abundance $\Omega_\ax h^2$ given its initial field value $\phi_\ini$ for $\Omega_\ax h^2\rightarrow 0$. Numerical results from \axie (points, blue solid curve) are plotted against the analytic fitting form (Eq.~(\ref{eq:Ffit}), gold dashed line). In the intermediate mass regime $m_\ax\sim H_\eq$, this step function form is fitted to numerical results.  In the $m_\ax \sim H_0$ dark energy regime, it reflects the time average of the oscillating density. 
    }
    \label{fig:F_and_fit}
\end{figure}

In Fig.~\ref{fig:F_and_fit}, we show the comparison between the numerical result and the analytic fitting function $F$.
Note that in the $m_\ax \sim H_0$ regime between matter and $\Lambda$ domination, the smooth transition is meant to match the time-averaged abundance rather than the instantaneous oscillatory abundance even when the \axie results report the latter ($m_\ax \le 10 H_0$,
independently of the setting of $m_\ax/H_*$). In the $m_\ax\gg H_0$ regime the fit is accurate at the sub-percent level.

This abundance relation also sets the amplitude of the initial isocurvature density power spectrum $A_\iso$ in terms of $r_\ax$.
Solving Eq.~(\ref{eq:F}) for $\phi_\ini$  and using $\delta_{\ax}=\delta\rho_\ax/\rho_\ax =2\delta\phi_\ini/\phi_\ini$ and Eq.~(\ref{eq:fielddimspec}), we obtain 
\begin{align}
\label{eq:AtoR}
A_{\iso}\equiv{}&
\frac{k^3 P_{\delta_\ax}}{2\pi^2}(k_0,\tau=0)\nonumber\\
={}&\frac{4}{\phi_{\rm ini}^{2}}\Delta_{\delta \phi_{\ini}}^{2}(k_0) =\frac{A_{s}r_\ax}{2}\frac{\Mpl^{2}}{\phi_{\rm ini}^{2}},
\end{align}
and find that
\begin{align}\label{eq:AisoOverAs}
\frac{A_{\rm iso}}{A_s} & 
\approx \frac{ r_\ax}{2 \Omega_\ax}  \left( \frac{m_\ax}{H_0} \right)^2  
F\left[\lg\left( \frac{m_\ax}{H_\eq}\right)\right]
a_\osc^3.
\end{align}
Here and throughout, when $\tau$ is used in the context of time, it denotes conformal time $d\tau= dt/a$ and should not be confused with $\tau$ as a cosmological parameter, the Thomson optical depth through reionization.

\begin{figure}
    \centering
    \includegraphics[width=1\linewidth]{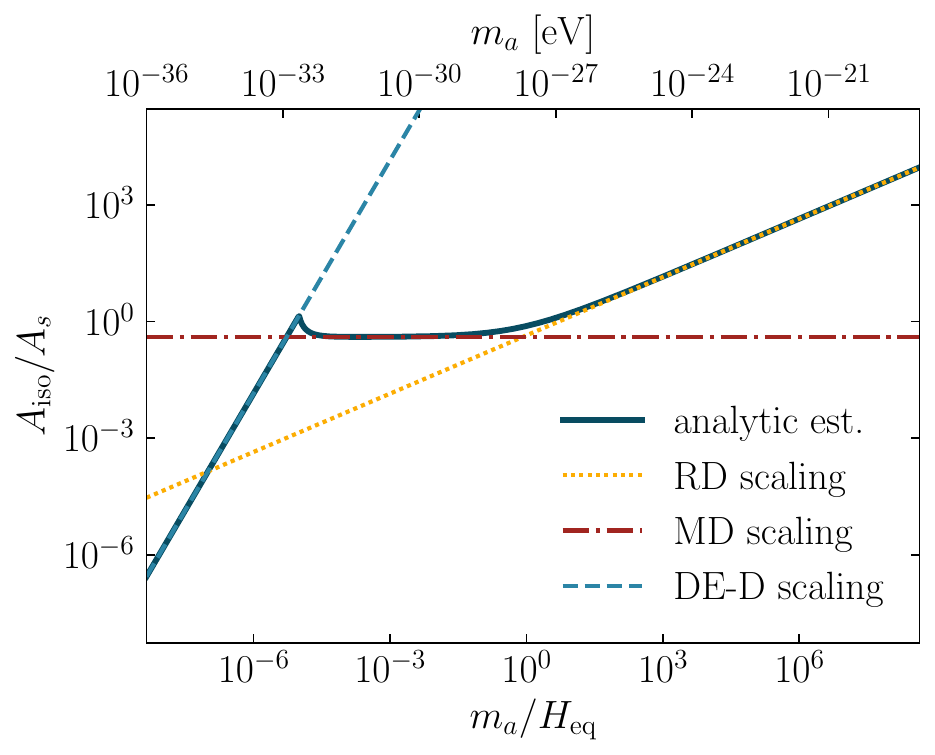}
    \caption{Analytic estimate for $A_\iso/A_s$ from Eq.~(\ref{eq:AisoOverAs}) as a function of $m_\ax$ at fixed $r_\ax = 0.036$ and $\Omega_\ax h^2 = 0.0024$. The analytic estimate joins the three asymptotic scaling cases where axion oscillations begin in the radiation dominated (RD), matter dominated (MD) and dark energy dominated (DE-D)  regimes.}
\label{fig:Aisofixedr}
\end{figure}

We plot this relation for $A_\iso/A_s$ in Fig.~\ref{fig:Aisofixedr}.
For definiteness we have taken $r_\ax=0.036$, which coincides with the current BICEP bounds on the tensor-to-scalar ratio and $\Omega_\ax h^2=0.0024$; the isocurvature amplitude would scale linearly with $r_\ax/\Omega_\ax h^2$ for other values.
Due to the mass scalings in the various regimes for $a_\osc$, shown with dashed lines, this isocurvature amplitude increases with increasing mass and is only flat in the matter dominated regime.   
The small feature at $m_\ax\sim H_0$ is due to the matching parameter $C=0.2$ which attempts to follow the time averaged definition of $\Omega_\ax h^2$ and not a physical effect. 

With these initial conditions for the field, the generalization of the 
\axie method for perturbations to isocurvature initial conditions is straightforward. All of the dynamical evolution of perturbations, detailed in Ref.~\cite{Liu:2024yne}, remains the same as for adiabatic modes with the change being that in the initial conditions only the axion field fluctuates.   
We set the conformal time derivative of the initial field fluctuation
\begin{equation}\label{eqn:perturbedIC}
\dot{\delta{\phi}}_\ini  \approx -\left(\frac{k^2}{3aH} + \frac{am_\ax^2}{5H}\right)\delta\phi_\ini.
\end{equation}
Here $\dot{\delta\phi}_\ini$ solves the KG equation as a perturbative series in $k/aH\ll 1$ and $m_\ax/H\ll 1$ at the initial time when the expansion is assumed to be radiation dominated $H \propto a^{-2}$. 

In \axie v1.1 we impose $m_\ax/H_{\rm ini}\lesssim 0.1$ as an additional criterion, though it has negligible effect for adiabatic conditions where field fluctuations are dynamically generated by curvature fluctuations.
Note that the usual power law expansion to higher powers in $k/aH$ (e.g. Ref.~\cite{Hlozek:2017zzf}) is unnecessary  as long as numerical mode evolution is started early enough for observable modes, whereas the correction in $m_\ax/H$ is important for all modes.
In fact, the background $\dot\phi_\ini$ solves the same equation with $k=0$ \cite{Liu:2024yne}.

CMB isocurvature power spectra are then computed by integrating the isocurvature radiation transfer functions against the initial axion power spectrum instead of the initial curvature power spectrum. Density power spectra are similarly constructed from the isocurvature transfer functions defined as
\begin{equation}\label{eq:T_i_iso}
T_i^{\rm iso} (k,\tau) =\frac{ \delta_{i}(k,\tau)}
{\delta_\ax(k,\tau_\ini)},
\end{equation}
where $i \in \ax,c,b,\nu$.  For $m_\ax/H_0>10$, all of these components are combined with their respective density weights into the matter power spectrum whereas for $m_\ax/H_0\le 10$ the axions are excluded from the sum.
For the axions the transfer function is defined to use the Klein-Gordon solution before the switch and the fluid after the switch at $\tau_*$. 
Density power spectra for the various species are then  given by
\begin{equation}
\label{eq:Piso}
\frac{k^3 P_{\delta_i}}{2\pi^2}(k,\tau) = A_{\rm iso} 
\left( \frac{k}{k_0} \right)^{n_\ax} [T_i^{\rm iso} (k,\tau)]^2.
\end{equation}
We also use $S_8=\sigma_8 (\Omega_m/0.3)^{1/2}$ as a monitor of the impact of axions on the matter power spectrum.   Here $\sigma_8$ is the rms amplitude of linear fluctuations in the matter density convolved with a spherical top hat filter of $8 h^{-1}$\,Mpc. 

Eqs.~(\ref{eq:F})-(\ref{eq:Piso}) fully specify the abundance, primordial isocurvature amplitude, and transfer-function normalization required to compute observable signatures. We now examine these signatures.

\section{CMB and Matter Power Spectra}\label{sec:cmb}

The observable signatures of axion isocurvature perturbations depend primarily on when the axion field begins to oscillate. This timing divides the phenomenology into three qualitatively distinct mass regimes. For $m_\ax \gg H_\eq$, the field begins oscillating before matter-radiation equality. In this regime, axions behave as cold dark matter on cosmological scales and their isocurvature signatures are nearly degenerate with those of conventional CDM isocurvature modes. For $H_0 \ll m_\ax \lesssim H_{\rm eq}$, Jeans stability in the matter dominated regime suppresses the growth of small-scale perturbations and breaks this degeneracy, leading to a characteristic scale-dependent cutoff in the CMB and matter power spectra. Finally, for $m_\ax \lesssim H_0$, the axion field has not yet begun coherent oscillations and instead behaves as dark energy, producing isocurvature signatures that are confined almost entirely to the largest angular scales through the late integrated Sachs-Wolfe effect.

In this section we characterize the CMB temperature and matter power spectra across these three regimes, derive analytic scaling relations that explain their behavior, and translate current and prospective observational constraints into bounds on the inflationary parameters that control axion isocurvature perturbations.

\begin{figure}
    \centering
    \includegraphics[width=1\linewidth]{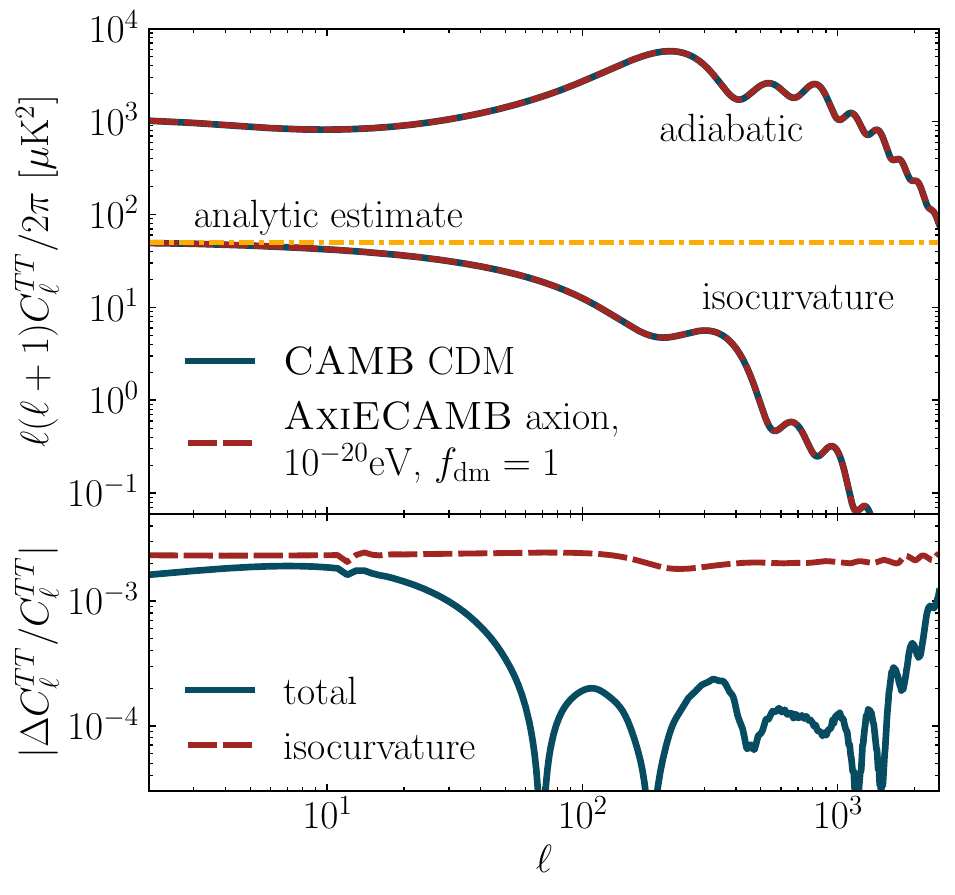}
    \caption{CMB temperature power spectrum $C_\ell^{TT}$ for both adiabatic and isocurvature components for axion dark matter compared with cold dark matter.  Here $m_\ax = 10^{-20}\,{\rm eV}\gg H_\eq$, $f_\dm = 1$ and $\beta_\iso = 0.03$ ($A_\iso\approx 6.5\times 10^{-11}$). The horizontal dashdot line marks the analytic estimate for the asymptotic low-$\ell$ behavior, calibrated at the quadrupole with $C_P=0.32$.  Bottom: the fractional difference between \axie and \camb for isocurvature and total (adiabatic + isocurvature) $C_\ell^{TT}$. For this $m_\ax \gg H_\eq$, axion isocurvature is observationally indistinguishable from conventional CDM isocurvature. 
    }
\label{fig:CAMB_CDM_compare}
\end{figure}
\begin{figure}
    \centering
    \includegraphics[width=1\linewidth]{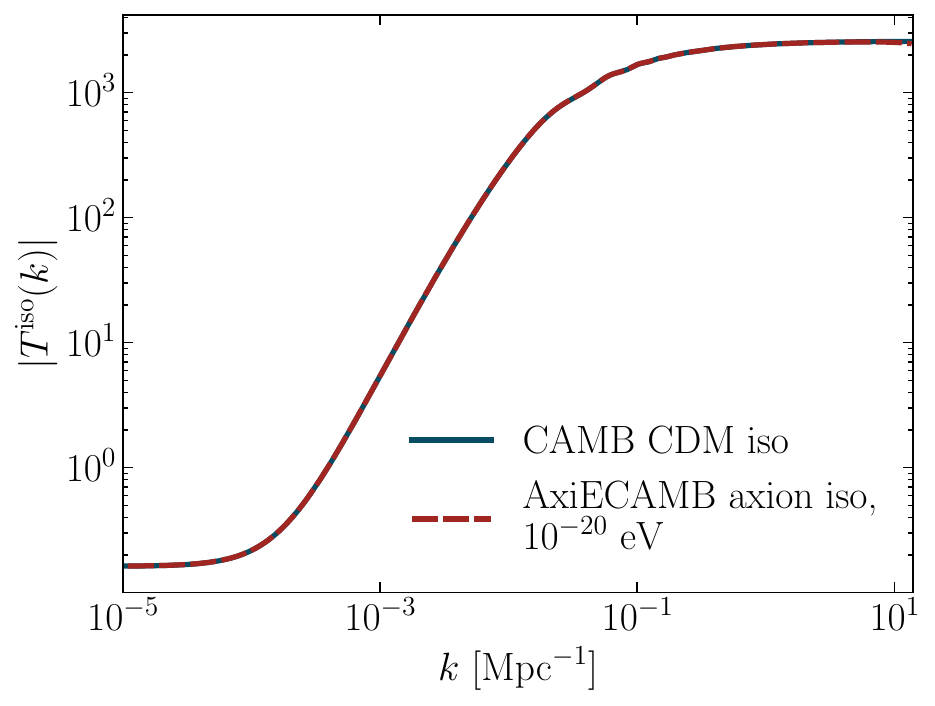}
    \caption{Axion isocurvature transfer function $T_\ax^\iso$  computed with \axie for a test $m_\ax =10^{-20}$\,eV with $f_\dm=1$ compared with the CDM isocurvature transfer function.  In this $m_\ax \gg H_\eq$ regime the two are indistinguishable, unlike the adiabatic $T(k)$. Here the raw \camb output has been converted to our transfer function convention (see Eq.~(\ref{eq:T_i_iso})).}
    \label{fig:CAMB_CDM_compare_Tiso}
\end{figure}

\subsection{Dark Matter Range}

Axions with $m_\ax\gg H_\eq \sim 10^{-28}$\,eV behave as CDM above their Jeans scale and thus the CMB observables for axions in this mass regime are equivalent to those of CDM. This is verified in Fig.~\ref{fig:CAMB_CDM_compare}, where we compare $C_\ell^{TT}$ for CDM isocurvature computed with \camb and axion isocurvature computed with \axie for $f_\dm = 1$.

In contrast to the usual adiabatic modes where the power spectrum rises with multipole $\ell$ from the low-$\ell$ Sachs-Wolfe plateau, the isocurvature spectrum falls.  When combined into the total spectrum, observational constraints on $r_\ax$ therefore mainly come from the already nearly cosmic-variance-limited range measured by Planck.  
This difference in the shape of the isocurvature power spectrum from the adiabatic one comes from the difference in the gravitational potential evolution. 
For modes that enter the horizon after matter-radiation equality ($k/aH|_\eq \ll 1$), the Newtonian gravitational potential $\Psi$ grows from zero linearly with the scale factor during radiation domination when $m_\ax \gg H$.   It reaches  
\begin{equation} 
\Psi \approx  - \frac{1}{5}\delta_m (\tau_\ini)
\label{eq:psiCDMiso}
\end{equation} during matter domination
and imparts an early ISW effect to the CMB of
$|\Delta T/T| \approx |2\Delta\Psi|$ 
[see~\cite{Hu:1994jd} Eq.~(27)].
We can therefore use this scaling to calibrate the low-$\ell$ plateau ``$P$'' for $\ell<\ell_\eq$ as  
\begin{align}\label{eqn:CP_for_plateau}
\frac{\ell(\ell+1)C_\ell^{TT}}{2\pi}\Big|_{\ell=2} ={}&  f_\dm^2 A_{\rm iso} C_P^2 \\
={}& \frac{1}{2} f_\dm^2 A_{s} r_\ax \left( \frac{\phi_\ini}{\Mpl} \right)^{-2} C_P^2,
\nonumber
\end{align}
where the suppression of $f_{\rm dm}^{2}$ arises when axions only contribute a fraction $f_{\rm dm}$ of the dark matter density fluctuation.

Numerical evaluation with \axie gives $C_P= 0.32$ instead of the analytic estimate given in Eq.~(\ref{eq:psiCDMiso}) of $0.4  \Omega_\dm/(\Omega_\dm+\Omega_b)\approx 0.34$.  We choose $\ell=2$ to calibrate the plateau even though it receives contributions from the late ISW effect from dark energy, since that is a much smaller relative effect for isocurvature modes than adiabatic modes (see
 \cite{Hu:1994jd}, Fig.~7).

For $k/aH|_\eq \gg 1$, or equivalently $\ell \gg \ell_\eq$, the gravitational potential ceases to grow linearly with $a$ at horizon crossing where $k=aH$ with $aH \propto a^{-1}$ in the radiation dominated epoch and the corresponding modes are suppressed as 
$(aH/k)|_\eq$ [see e.g.~\cite{Hu:1994jd} Eq.~(43)], leading to an additional suppression of  $\ell^{-2}$ in power.  This explains the suppression of isocurvature contributions relative to adiabatic ones in the acoustic regime.

We also compare the axion transfer function to the CDM isocurvature transfer function in Fig.~\ref{fig:CAMB_CDM_compare_Tiso}.  Note that for $k \gg k_\eq $, the axion density fluctuation grows from its initial value after equality in a scale-free manner $\propto f_\dm a/a_{\rm eq}$ whereas for $k \lesssim k_\eq$ the scaling reflects instead the scale-free gravitational potential amplitude from Eq.~(\ref{eq:psiCDMiso}) and hence $(k/k_\eq)^2$ scale-dependent suppression from the Poisson equation.  For $k\ll k_\eq$ it again reaches a scale-free but suppressed level.  The suppression reflects 
the decay of the initial conditions until the point where the induced fluctuations in the other energy-density components can compensate so as to maintain the total isocurvature condition $\sum \delta\rho_{i}=0$ to leading order.  In the matter dominated epoch this compensation is between matter-like species so that
\begin{equation}
T_{\ax}^{\iso}(k\rightarrow 0)\sim \frac{\Omega_b + \Omega_c}{\Omega_b+\Omega_\dm} \le 1.
\label{eq:axionCIP}
\end{equation} 
This is analogous to a compensated isocurvature fluctuation, where the compensation of the axion fluctuation is in both the baryons and the cold dark matter.

Given the agreement between predictions for axion and CDM isocurvature power spectra in the relevant regimes, observational bounds on CDM isocurvature can be directly translated to  axion isocurvature bounds. For the Planck constraint on uncorrelated, scale-invariant CDM isocurvature from TT,TE,EE+lowE and CMB lensing power spectra \cite{Planck:2018jri} (see also \cite{Petretti:2026ayw}),
\begin{equation}
\beta_{\rm iso} < 0.038, \quad (95\% \, {\rm CL})
\end{equation}
where
\begin{align}
\beta_\iso ={}& \frac{A_\iso f_\dm^2}{A_\iso f_\dm^2+A_s } 
\nonumber\\
={}& \left[  1 + \frac{2}{r_\ax} \left( \frac{\phi_\ini}{\Mpl f_\dm}  \right)^2 \right]^{-1}.
\label{eq:betaiso}
\end{align}

The Planck constraint can then be converted to a bound on $r_\ax$.
Using Eq.~(\ref{eq:abundanceRD}) and (\ref{eq:betaiso}) we obtain
\begin{align}
\left(\frac{\phi_\ini}{\Mpl}\right)^2 \approx 2.41 \frac{\Omega_\ax }{\Omega_\dm +\Omega_b } 
\left( \frac{m_\ax}{H_\eq} \right)^{-1/2},
\end{align}
and thus
\begin{align}\label{eq:r_a_bound_RD}
r_\ax \approx{}& \frac{4.8}{f_\dm} \frac{\beta_{\rm iso}}{1-\beta_{\rm iso}} \frac{\Omega_\dm}{\Omega_b+\Omega_\dm}
\left( \frac{H_\eq}{m_\ax} \right)^{1/2}
\nonumber\\
<{}&
\frac{0.076}{f_\dm}
\frac{\Omega_\dm h^2}{0.12}
\left( \frac{m_\ax}{10^{-27}\,{\rm eV}} \right)^{-1/2},
\end{align}
where we have assumed $m_\ax\gg H_\eq$ and note that the constraint on $\beta_\iso$ from Planck requires compatibility with its $\Omega_\dm h^2$ measurement from adiabatic modes.\footnote{With an axion mass of $m_\ax=10^{-24}~{\rm eV}$, this yields $r_\ax< 0.0024 f_\dm^{-1}$, in contrast to  Ref.~\cite{Hlozek:2017zzf} where Markov Chain Monte Carlo limits of $r_\ax \lesssim 10^{-2}$ at $f_\dm =1$ are reported (their Fig.~10; see Appendix \ref{app:accuracy} in this work for the improvement due to our more accurate treatment of the abundance).  The analytic expressions there  [Eq.~(12)] have the same scaling with $f_\dm$ but were improperly normalized by a large factor, which cannot be explained simply by the less precise treatment of $F_{0}$ in that work.
This scaling is furthermore not reproduced in Fig.~10 of \cite{Hlozek:2017zzf}. While this is conceivably a result of marginalization over other parameters with their priors, there are no obvious degeneracies responsible for this fact.  We do not explore this issue further.}

We can extend these $m_\ax \gg H_\eq$ relations to $m_\ax \sim H_\eq$ with the help of the more general abundance formula Eq.~(\ref{eq:F}) to convert $\beta_\iso$ into $r_\ax$ through $\phi_\ini$.   This formula is accurate at the $1\%$ level for all $f_\dm$ when $m_\ax>10^{-27}$\,eV  and at the 10\% level for $m_\ax=10^{-28}$\,eV and $f_\dm=1$ with the accuracy rapidly improving for lower $f_\dm$ at this mass.

\begin{figure}
    \centering
    \includegraphics[width=1.\linewidth]{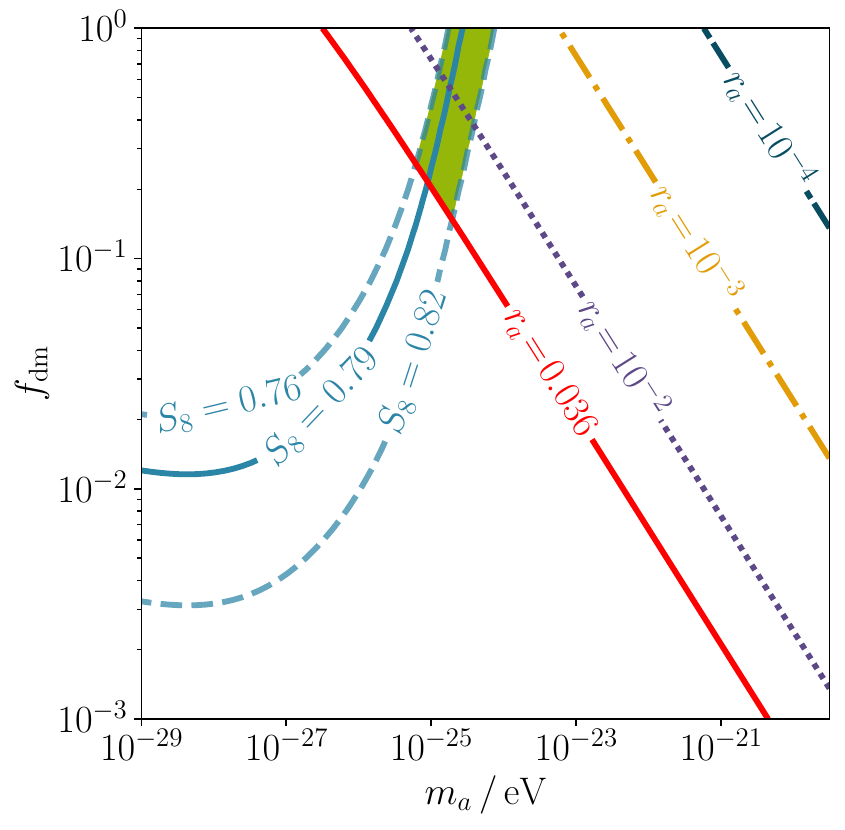}
    \caption{
 Upper limit (95\% CL diagonal lines) on  tensor-to-scalar ratio $r_\ax$ from the Planck 2018 isocurvature bound in the $r_\ax<0.036$ regime allowed by the current BICEP constraints ($r_\ax < 0.036$, red line).  The green-shaded region marks the intersection of the observationally interesting $S_8=0.76$--$0.82$ band (blue curves) and identifies a window of coexistence where both inflationary isocurvature and tensors could be discovered in the near future. Detection of isocurvature in the $r_\ax>0.036$ lower-left region would imply isocurvature generation beyond this minimal scenario. }
    \label{fig:rbounds_and_S8}
\end{figure}

This isocurvature bound should be compared to the bound on $r$ from the absence of gravitational wave B-modes from BICEP/Keck
\cite{BICEP:2021xfz}, $r_{0.05}<0.036$ (95\% CL).   With Eq.~(\ref{eq:F}) and $r_\ax=r$, the Planck isocurvature bound is more powerful than the tensor bound if
\begin{equation}
m_\ax > 
\begin{cases}
3 \times 10^{-27} \,{\rm eV} & f_\dm =1 \\
4 \times 10^{-27} f_\dm^{-2}\,{\rm eV} & f_\dm \ll 1
\end{cases}
\end{equation}
The slight deviation from the power law $f_\dm^{-2}$ near $f_\dm=1$ is due to the mass limit approaching $H_{\eq}$ for the current BICEP constraint where the abundance formula transitions in Fig.~\ref{fig:F_and_fit}.  The deviation would disappear 
for future stronger bounds on $r$.
In Fig.~\ref{fig:rbounds_and_S8},  we show the axion parameters $m_\ax,f_\dm$ where the upper limit on $r_\ax$  from the isocurvature bound of $\beta_\iso<0.038$ takes on values that are stronger than the current BICEP bound.

Unlike the BICEP bound which relies on the non-detection of tensor modes, for an isocurvature bound on the tensor-to-scalar ratio to apply, there must be a detection of axions themselves with a determination of $m_\ax$ and $\Omega_\ax h^2$. For example, suppose that a future constraint on the weak lensing parameter $S_8$ implies a suppression relative to the $\Lambda$CDM prediction at high significance.  Since axions suppress power in the adiabatic mode for $k>k_{\rm J}$, this would indicate a preference for axions of a certain mass and abundance.  

In Fig.~\ref{fig:rbounds_and_S8}, we show the effect of axions on $S_8$ over the observationally relevant range for the current, low significance, tension between the CMB and weak lensing. For example, the DES Y6 weak lensing measurement implies $S_8 =0.798^{+0.014}_{-0.015}$ \cite{2026arXiv260210065D}.
We highlight the interesting region in green for a significant lowering of $S_8$ from the CDM value in this model of $S_8={0.83}$, which can satisfy both the current BICEP and isocurvature bounds.

In this region, a near-future detection of isocurvature modes would predict a near-future target for tensor modes and vice versa.
Conversely, a null detection of isocurvature modes could be used to strengthen the bounds on inflation in this region.

\subsection{Intermediate Range}

In the intermediate mass range 
$10^{-32} \lesssim m_\ax/\text{eV}\lesssim  10^{-28}$, the impact of axions on adiabatic modes leads to strong bounds on $f_{\rm dm}$ so as not to 
 adversely impact the CMB and large scale structure.
We can see this in Fig.~\ref{fig:rbounds_and_S8} from the existence of a region where the weak lensing parameter $S_8 \sim 0.76$--$0.82$, and is not overly suppressed from its $\Lambda$CDM value of $S_8=
0.83$.  Similarly, CMB constraints  require $f_\dm < 0.05$ at 95\% CL \cite{Hlozek:2014lca}, compatible with the $S_8$ band.  In terms of amplitude, the impact of axion isocurvature modes relative to adiabatic modes thus becomes less observable.

On the other hand, in this regime, axion isocurvature modes are no longer degenerate in shape with CDM isocurvature modes for the CMB.
This degeneracy breaking is related to the Jeans stability of the axions on small scales.
While the low $\ell$ plateau in TT remains relatively unchanged from its CDM form, the cutoff moves to lower multipoles because the effective Jeans scale of axion perturbations exceeds the horizon at equality.

Both the plateau and cutoff in the TT phenomenology are determined by the sound speed of the axions.
The effective sound speed of pressure waves is \cite{Passaglia:2022bcr}
\begin{align}
\label{eq:cs2}
c_\ax ={}&    \frac{ \sqrt{1+\left({k}/{a m_\ax} \right)^2} -1}{{k}/{am_\ax}},
\end{align}
which defines the sound horizon or Jeans scale as $\int d\tau c_\ax$. After the beginning of oscillations, the sound speed drops from unity and so superhorizon scale modes at $a_\osc$ are always above the sound horizon thereafter.
For these modes, the gravitational potential $\Psi$ again grows to $\Psi \sim -\delta_m(\tau_\ini)/5$ for $a\gg a_\osc$.  

The qualitative reason is that in the presence of  a long wavelength axion perturbation, the local expansion history  mimics a separate universe background evolution where the axion abundance is locally different than the true background.  While the growing local curvature  technically violates an exact Friedmann-Robertson-Walker assumption, its impact on the local Hubble rate is suppressed as $(k/aH)^2$ \cite{Hu:2016ssz}.

After $a \gg a_\osc$ in this local patch the dark matter density differs from the background in the same way as it would be for CDM isocurvature modes.   Since the CMB early ISW effect is sensitive mainly to the total change $2\Delta\Psi$, the amplitude is nearly the same. This equivalence also requires the true background expansion history to be close to $\Lambda$CDM, which it is for the allowed $f_\dm\ll 1$.

   \begin{figure}
        \centering
        \includegraphics[width=1\linewidth]{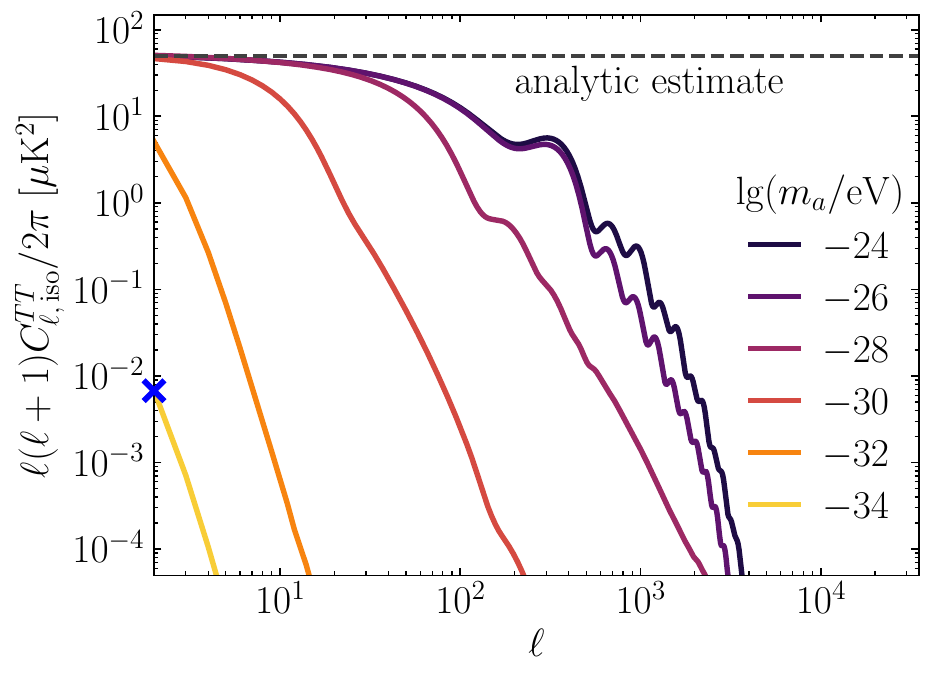}
        \caption{Isocurvature $C_\ell^{TT}$ as a function of $m_\ax$ with fixed $A_\iso=  1.6\times 10^{-7}$, $\Omega_\ax h^2=0.0024$, $\Omega_c h^2= 0.1176$ ($f_\dm = 0.02$ and $\beta_\iso = 0.03$ for $m_\ax>10 H_0$) and all other parameters held to their fiducial values. 
        Dashed horizontal line marks the analytic estimate for the  low-$\ell$ plateau  with $C_P = 0.32$ assuming $m_\ax\gg H_0$ (see Eq.~(\ref{eqn:CP_for_plateau})).
        Cross ($\times$) represents the $C_Q$ analytic estimate for the quadrupole for $m_\ax \ll H_0$. 
        The low-$\ell$ plateau is nearly mass-independent in the dark matter regime, becomes increasingly suppressed 
       for $H_0\ll m_\ax\ll H_{\rm eq}$, leaving only a signal at the first few multipoles for $m_\ax \lesssim H_0$. For $m_\ax \le 10^{-33}$\,eV, the $C_\ell^{TT}$ curves are almost indistinguishable.
        }\label{fig:fDM0d02_fixedbetaiso}
        
    \end{figure}
    
   \begin{figure}
        \centering
        \includegraphics[width=1\linewidth]{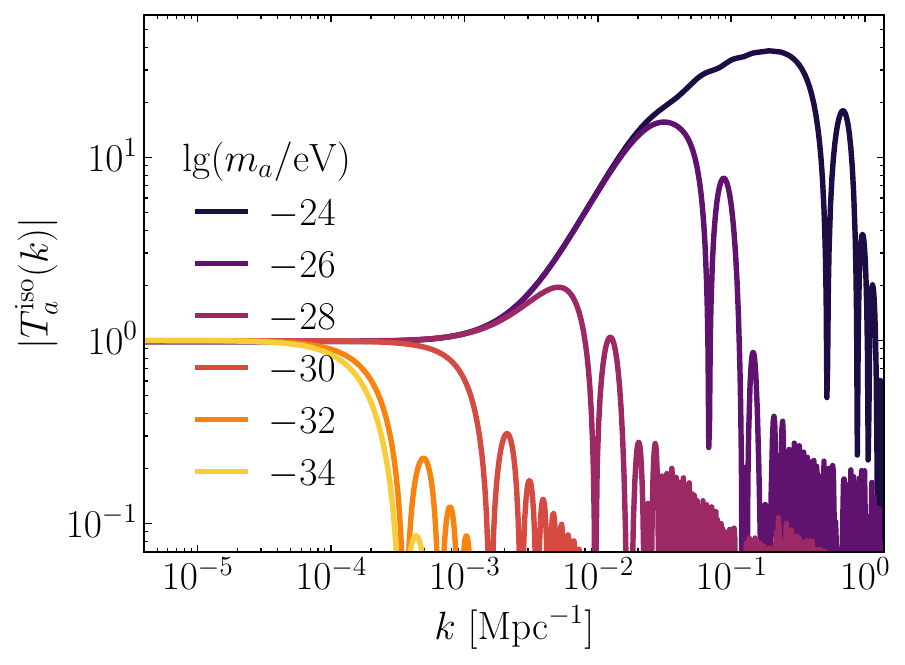}
        \caption{Axion isocurvature transfer function $T_\ax^\iso$ as a function of $m_\ax$ with fixed $A_\iso=  1.6\times 10^{-7}$, $\Omega_\ax h^2=0.0024$, $\Omega_c h^2= 0.1176$ ($f_\dm = 0.02$ and $\beta_\iso = 0.03$ for $m_\ax>10 H_0$).  Compared to the CDM isocurvature transfer function from Fig.~\ref{fig:CAMB_CDM_compare_Tiso}, the main change is the suppression and oscillation for $k\gg k_J$ and its shift to lower wavenumbers at lighter masses.  
        \label{fig:Tiso}
   }
   \end{figure}

On the other hand, isocurvature contributions to CMB anisotropy from scales smaller than the horizon at $a_\osc$, $k/a_\osc H = 3k/a_\osc m_\ax\sim 1$  are suppressed due to Jeans stability.   For $m_\ax<H_\eq$ the axion sound horizon ceases to grow at $a\gtrsim a_\osc$ since the sound speed decreases as $c_\ax \propto k/ am_\ax$.
The characteristic start of the suppression scales with the horizon $\ell_J \propto a H|_\osc \propto m_\ax^{1/3}$ for $H_0 \ll  m_\ax\ll H_\eq$.  As $m_\ax$ approaches $H_0$ and drops below it, the suppression scale saturates at the current horizon and becomes independent of $m_\ax$.  Furthermore, the $a^3$ growth in the synchronous-gauge curvature variable $\eta$
[see \cite{Hu:2016ssz} Eq.~(105)] ceases at horizon crossing ($a H \propto a^{-1/2}$) during matter domination. 
At this point the Jeans oscillations in the axions produce oscillations in the Newtonian potential and when combined into the ISW response lead to a suppression  in the temperature anisotropy power spectrum that scales as $\sim (\ell_J/\ell)^6$ for higher multipoles.

Fig.~\ref{fig:fDM0d02_fixedbetaiso} illustrates these effects on the isocurvature TT power spectrum across a wide range of masses $m_\ax/\text{eV}\in [10^{-34}, 10^{-24}]$. 
Here we adopt fixed parameters at $A_\iso=  1.6\times 10^{-7}$, $\Omega_\ax h^2=0.0024$, $\Omega_c h^2= 0.1176$
which for the $m\gg H_0$ cases is equivalent to $f_\dm=0.02, \beta_\iso=0.03$ whereas for the $m_\ax \lesssim H_0$ cases $f_\de=0.0077$, and fiducial $\Lambda$CDM parameters otherwise. 
Note that by fixing $A_\iso$ instead of $r_\ax$, we highlight that the phenomenological signatures of axion isocurvature modes hold beyond the specific context of their inflationary origin.

We include the analytic prediction for the low-$\ell$ plateau from Eq.~(\ref{eqn:CP_for_plateau}) with $C_P=0.32$. For the dark energy regime ($m_\ax \lesssim 10^{-33}$\,eV) discussed in the next section, the analytic prediction for the quadrupole is  marked with a cross.
For $H_0 \ll m_\ax \ll H_\eq$ the power at  $\ell\gg \ell_J\propto m_\ax^{1/3}$ also falls off in a manner consistent with the $\ell^{-6}$ analytic expectation for the suppression due to Jeans stability.

These considerations also determine the form of the axion transfer function. Below the Jeans scale, field fluctuations behave as radiation and oscillate with an amplitude that decays as $\delta\phi \propto a^{-1}$ in the absence of metric sources. Equivalently the kinetic energy density of a mode on its own also scales as radiation $\propto a^{-4}$.
While $m_\ax \ll H$, the field is nearly frozen and the background kinetic energy scales $\propto 1/H^2$ whereas the total background density is nearly constant. 
Thus the kinetic energy density fluctuation of the mode scales as  $\delta_\ax \propto 1/(a^2 H) \propto a^{(3w_T-1)/2}$.
Unlike the field fluctuations, the isocurvature density perturbations oscillate but do not decay in radiation domination and decay as $a^{-1/2}$ in matter domination.

The  resulting axion transfer function is shown in Fig.~\ref{fig:Tiso} for the same range in $m_\ax$.   For $m_\ax\gg H_\eq$, it follows the pattern of Fig.~\ref{fig:CAMB_CDM_compare_Tiso} until $k>k_J$.
Here the transfer function sharply falls. Then $\delta_\ax$ oscillates, but does not decay during radiation domination. There is thus a floor to the high $k$-suppression in $\delta_{a}$, due to the delay in growth after $a_\eq$ because of Jeans stability.
As $m_\ax$ drops below $H_\eq$, this cuts into the $(k/k_\eq)^2$ ramp up of the transfer function. The compensated tail as $k\rightarrow 0$ follows Eq.~(\ref{eq:axionCIP}) but we have chosen a small $\Omega_\ax/\Omega_\dm$ value here so that $T_\ax^{\iso}(0) \sim 1$ (cf.~Fig.~\ref{fig:CAMB_CDM_compare_Tiso}) where it takes a smaller value $\sim \Omega_b/(\Omega_b+\Omega_\ax)$.

In this regime well under the Jeans scale, axion density fluctuations do decay before $a_\osc$ leading to a power law suppression.  In the highly suppressed regime, the oscillations are too fast compared with the mass scale and the numerical results should only be used for order of magnitude estimates of the very high suppression relative to adiabatic modes (see Appendix \ref{app:ETA}).

This relative fall off of the anisotropy spectrum from a plateau set by essentially the same $C_P$ value as with CDM isocurvature modes means that axion isocurvature constraints from the CMB will degrade as $m_\ax$ decreases below $H_\eq$.
To obtain an analytic scaling, if we take the same definition of $\beta_{\rm iso}$ as in Eq.~(\ref{eq:betaiso}) but for $H_0\ll m_\ax \ll H_\eq$ where Eq.~(\ref{eq:F0}) applies, then
\begin{equation}
r_\ax = \frac{16}{3} \frac{\beta_{\rm iso}}{1-\beta_{\rm iso} }\frac{\Omega_\dm}{f_\dm \Omega_m}.
\end{equation}
For the cases in Fig.~\ref{fig:fDM0d02_fixedbetaiso}, this gives
$r_\ax\approx 6.9$ which is clearly already ruled out by constraints on tensor modes.  Note that this analytic estimate uses the $f_\dm\ll 1$ abundance formula.  As discussed in Appendix \ref{app:accuracy}, at the worst case of $f_\dm =1$, which is already ruled out, this is about $10\%$ accurate for  $m_\ax=10^{-28}$\,eV and climbs to $\lesssim {\cal O}(1)$ at even lighter masses. For allowed $f_\dm$, it is accurate at $\lesssim$ 1\% across the entire DM regime $m_\ax \gg H_0$.

Conversely, the BICEP bound $r_\ax<0.036$ places an upper bound on the isocurvature amplitude for the allowed $f_\dm\ll 1$ regime of
\begin{equation}
\beta_\iso < 0.0002 \frac{f_\dm}{0.03} \frac{\Omega_m}{\Omega_\dm},\qquad (r_\ax<0.036)
\end{equation}
or
\begin{align}
\frac{A_\iso}{A_s} ={}& \frac{3}{16} \frac{\Omega_m}{\Omega_\dm}\frac{r_\ax}{f_\dm}\nonumber\\
<{} & 0.23 \frac{\Omega_m}{\Omega_\dm} 
\frac{0.03}{f_\dm}, \qquad (r_\ax<0.036).
\end{align}
Note that as $f_\dm \rightarrow 0$, $A_\iso\gg A_s$ is allowed since its observational effect also goes to zero.

\begin{figure}
    \centering
    \includegraphics[width=1\linewidth]{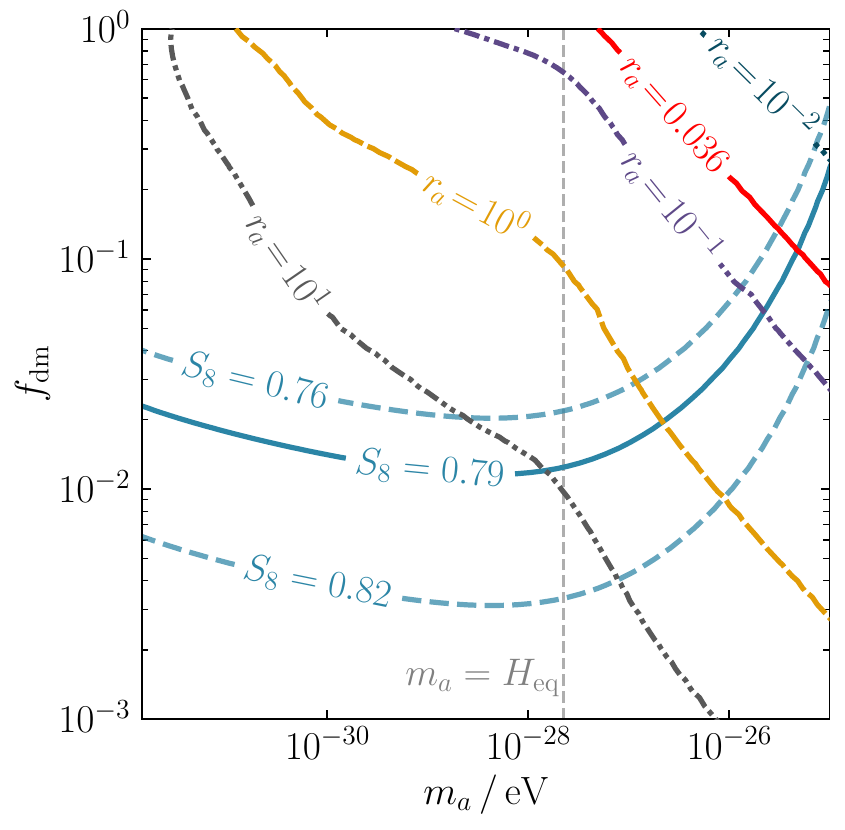}
    \caption{Fisher forecast of the $95\%$ CL upper limit on $r_\ax$ from isocurvature modes.  Here we intentionally take overly optimistic assumptions: cosmic-variance limited measurements to $\ell\le2000$ and no  degeneracies between other parameters and $r_\ax$.  For $m_\ax\gg H_\eq$ the forecasted bounds are a factor of 2.6 stronger than current bounds in Fig.~\ref{fig:rbounds_and_S8}.    For $m_\ax \ll H_\eq$, even this optimistic forecast provides weaker bounds than the current ones from tensor modes $r_\ax<0.036$ (red solid line) implying that inflationary isocurvature modes are unobservable by these means.
    The blue curves again show the observationally interesting $S_8=0.76$--$0.82$
    region.
    }
    \label{fig:Fisher_contours_DM}
\end{figure}

Even if measurements of the CMB primary anisotropy improve to their cosmic variance limit, isocurvature bounds on $r_\ax$ in this range of $m_\ax,f_\dm$ cannot become stronger than the tensor bounds.  We demonstrate that with a Fisher matrix forecast which takes intentionally optimistic assumptions: all parameters but $r_\ax$ are fixed such that any degeneracies between parameters must be resolved by other means, and the covariance of $X,Y\in TT,TE,EE$ measurements is cosmic variance limited with $f_{\rm sky}=1$ over $2\le\ell\le\ell_{\rm max}=2000$. The forecasted inverse variance of $r_\ax$ is then 
\begin{equation}
\label{eq:Fisher}
\sigma_{r_\ax}^{-2}=F_{r_\ax r_\ax}=\sum_{\ell,X,Y} \frac{\partial C_\ell^{X}}{\partial r_\ax}
{\rm Cov}_{XY}^{-1} \frac{\partial C_\ell^{Y}}{\partial r_\ax},
\end{equation}
where $X,Y\in TT,TE,EE$, and ${\rm Cov}_{XY}$ is the Gaussian covariance built from the fiducial total spectra. We evaluate the Fisher at the fiducial point $r_\ax=0$.
We show the forecasted $95\%$ upper limit on $r_\ax$ in Fig.~\ref{fig:Fisher_contours_DM}, using the Fisher $2\sigma$ value as a proxy
for the one-sided upper limit bounded at $r_\ax=0$.
At $m_\ax\gg H_\eq$ they imply $\beta_\iso<0.0145$, a factor of $2.6$ improvement over the current Planck bound.

For $m_\ax \sim H_\eq$, the upper limits on $r_\ax$ become less dependent on $m_\ax$ since the relationship between $\Omega_\ax h^2$ and $\phi_\ini$ becomes independent of mass.  On the other hand for $m_\ax \ll H_\eq$ the constraints further weaken because the Jeans cut off $\ell_J$ suppresses the isocurvature signal. Even with this optimistic forecast, isocurvature modes from inflationary models that are consistent with current tensor constraints are undetectable for $m_\ax< H_\eq$.

\subsection{Dark Energy Range}

The phenomenology of axion isocurvature modes changes qualitatively for masses where the background field has not yet entered into its rapid oscillation stage  by the present. 
For the mass range $m_\ax \lesssim 10 H_0$, axions behave like part or all of the dark energy.   Recent studies have shown that the $f_\de$ values which best fit current CMB, BAO and SN data rise rapidly from $\sim 0.1$ at $m_\ax\sim 10^{-32.5}$\,eV to $1$ at $m_\ax \lesssim 10^{-32.9}$\,eV 
\cite{Liu:2025bss}.

In this mass range, axion isocurvature fluctuations produce a late-time ISW contribution to CMB temperature anisotropy.   These contributions peak at the quadrupole and fall sharply to higher multipoles \cite{Gordon:2004ez}
\begin{align}
\label{eq:dequadrupole}
\frac{{\ell}({\ell+1})C_\ell^{TT}}{2\pi}\Big|_{\ell =2}
={}&
\frac{3}{\pi} C_2^{TT} =f_\de^2  A_{\rm iso}  C_Q^2,
\end{align}
where the scaling coefficient $C_Q \approx (1/75)\allowbreak(\Omega_\de/0.73)\allowbreak (\Omega_m/0.27)^{-1} \approx 0.011$ 
in Ref.~\cite{Gordon:2004ez} (their Eq.~(22) combining the 1/15 suppression with the adiabatic quadrupole factor 1/5). 

This suppression comes largely from the fact that $\delta_{a}$ modes that are inside the horizon before dark energy domination have decayed and contribute negligibly. Additionally, super-horizon modes contribute (via the ISW effect)  mainly to the unobservable monopole, leaving the quadrupole contributions suppressed by an order of magnitude. We use \axie to provide a more accurate calibration in the limit $f_\de=1$ and $m_\ax \ll H_0$ and find
\begin{equation}
C_Q= 0.0097
\end{equation}
for the fiducial parameters. 
For $\ell \gg 2$ the same $\ell^{-6}$ relative falloff as with higher masses occurs in the dark energy regime (see Fig.~\ref{fig:fDM0d02_fixedbetaiso}). As shown in  Appendix \ref{app:accuracy}, this analytic estimate is a good $\sim 10\%$ approximation or better for $m_\ax\lesssim 10^{-33}$\,eV for any $f_\de$. 

Using Eq.~(\ref{eq:abundanceDED}), for a fixed $r_\ax$, the isocurvature contribution to $C_\ell^{TT}$ scales as $(m_\ax/H_0)^2$ for $m_\ax \ll H_0$  and
\begin{equation}
 \frac{3}{\pi} C_2^{TT}  \approx  \frac{C_Q^2}{12} \frac{\Omega_\ax}{\Omega_{\rm de}^2} 
\left( \frac{m_\ax}{H_0} \right)^2 {A_s r_\ax  } .
\end{equation}
Thus the BICEP bounds on $r_\ax=r<0.036$ require a highly suppressed isocurvature contribution in this regime. 

\begin{figure}
    \centering
    \includegraphics[width=1\linewidth]{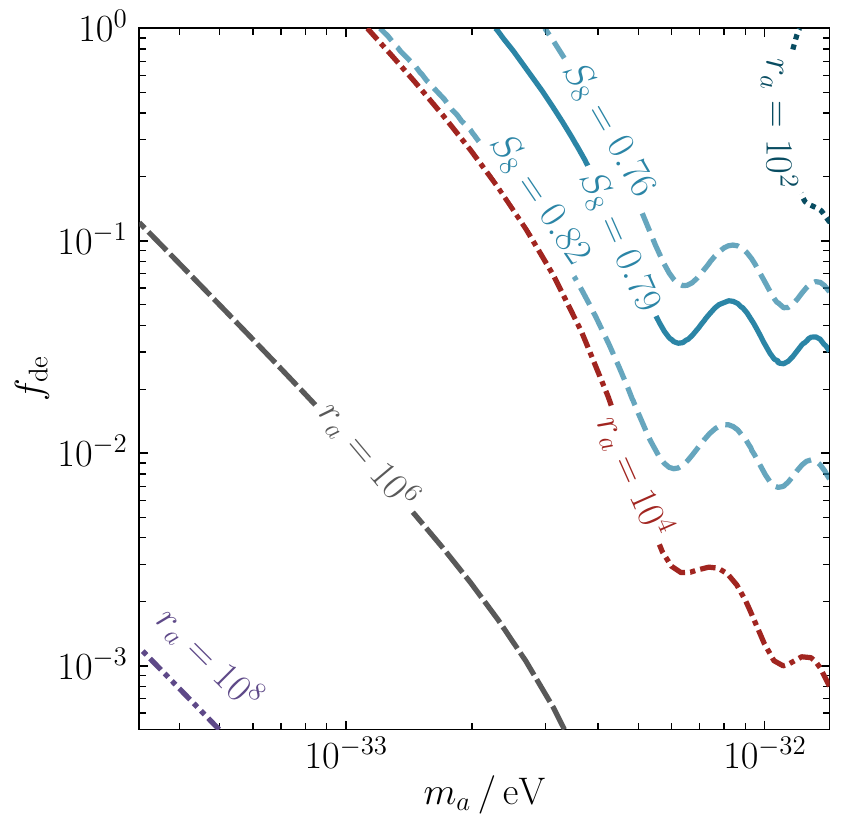}
    \caption{
    Fisher forecast of the $95\%$ CL upper limit on the isocurvature $r_\ax$ as in Fig.~\ref{fig:Fisher_contours_DM}, but in the DE-like axion mass range. The oscillatory structure at $r_\ax = 10^4$ near $m_\ax\sim10^{-32}$\,eV reflects the onset of axion field
    oscillation in $C_\ell^\iso$ and is dominated by the signal at $\ell = 2$. The curved feature at $r_\ax=10^2$ is due to the interplay between enhanced ISW sources and their distance-modified projection onto the low multipole anisotropy.  In the whole parameter space, the suppression of isocurvature contributions and current tensor bounds means that no CMB detection of inflationary isocurvature modes is possible. The blue curves again show the observationally interesting $S_8=0.76$--$0.82$ region, whose $S_8\sim 0.82$ boundary closely tracks the favored observational region for CMB+BAO+SN measurements \cite{Liu:2025bss}.
    }
    \label{fig:Fisher_contours_DE}
\end{figure}

Conversely, the isocurvature upper limit on $r_\ax$ scales as $f_\de^{-1} m_\ax^{-2}$ and is exceedingly weak,
weaker than even the linearization constraint of $A_\iso<1$ for sufficiently small $f_\de$.  In Fig.~\ref{fig:Fisher_contours_DE}, we repeat the calculation of the optimistic Fisher bound from Eq.~(\ref{eq:Fisher}) for this mass regime.  In particular, for $m_\ax\ll H_0$ this gives
\begin{equation}
r_\ax < 1.3 \times 10^4 \, f_\de^{-1} \left( \frac{m_\ax}{10^{-33}\,{\rm eV}}\right)^{-2} \quad (95\% \,{\rm CL})
\end{equation}
Observable axion dark energy isocurvature modes in the CMB therefore require a production mechanism beyond the inflationary misalignment mechanism with an initially frozen field in a nearly quadratic potential \cite{Gordon:2004ez}. 

Unrelated to isocurvature bounds, it is still interesting to note that for the adiabatic modes, the same region of the $m_\ax$--$f_\de$ plane picked out by $S_8 \sim 0.82$ happens to closely coincide with the  region favored by 
a recent analysis of DESY5 supernovae distances $+$ DESI BAO $+$ CMB  \cite{Liu:2025bss} (see their Fig.~2).  As with the dark matter masses, this region presents an interesting discovery window for axions in the near future.

\section{Discussion}
\label{sec:discussion}

Inflationary fluctuations in axion fields provide an observational connection between cosmological observables and the energy scale of inflation in the same way as primordial gravitational waves.   Realizing this connection in practice requires accurate predictions for how primordial field fluctuations map into late-time observables over the many orders of magnitude in axion mass, from the current Hubble scale to more than 10 orders of magnitude above it.   

In this work, we refine the tools to accurately evolve axion isocurvature perturbations from their inflationary origin through to the present epoch using the effective time average approach of \axie for switching from solving the exact field equations to the approximate fluid description. 
This includes improved initial conditions, later switches, and a monitor of when the ETA fails to effectively time average the axion system. 
We then use \axie to gain both physical intuition and determine observational constraints.

We provide sub-percent fitting formulae for the axion abundance over the entire range $m_\ax \gg H_0$ for allowed models, useful for relic abundance studies independent of isocurvature modes.  This then provides an accurate analytic calibration of the isocurvature amplitude from the tensor-to-scalar ratio of inflation.  The resulting fitting formulae reproduce the numerical abundance calculation.  They exhibit simple scaling behavior in each asymptotic limit of the three mass regimes where the axions begin oscillating during radiation, matter, and dark-energy domination. These expressions provide a direct mapping between inflationary initial conditions, axion abundance, and observable isocurvature power.

Using these relations we map CDM isocurvature constraints onto axion isocurvature constraints for inflation, greatly refining numerical treatments in the existing literature and, for the analytic estimates in Ref.~\cite{Hlozek:2017zzf} shifting the relevant mass range by orders of magnitude.   We find that the Planck data constrain the product of the tensor-to-scalar ratio and axion fraction to $r_\ax f_\dm < 0.076 (m_\ax/10^{-27}\,{\rm eV})^{-1/2}$ for high masses in the dark matter regime, unless isocurvature modes are suppressed relative to the frozen-field case considered here \cite{Jeong:2013xta,Nomura:2015xil,Kawasaki:2015lpf,Takahashi:2015waa,Nakayama:2015pba,Graham:2025iwx,Harigaya:2015hha}.

Interestingly, there exists a compatibility window around $m_\ax \sim 10^{-25}$\,eV and $f_\dm \gtrsim 0.1$ where both axion isocurvature and tensor modes could yet be discovered while simultaneously alleviating the current $S_8$ tension in $\Lambda$CDM. 

Conversely, once current tensor bounds are imposed, inflationary axion isocurvature becomes unobservable in the primary CMB anisotropy for $m_\ax f_\dm^2 \lesssim 10^{-26.4}$\,eV.
This suppression of inflationary axion isocurvature in both the intermediate and dark-energy regimes implies that any future detection of these characteristic signatures in the primary CMB would point toward an origin that deviates from the standard inflationary misalignment mechanism with a frozen field on a nearly quadratic potential (e.g.\ \cite{Gordon:2004ez}).
Nontrivial axion dynamics that drive the field toward a region of enhanced sensitivity could enhance resulting isocurvature perturbations \cite{Kobayashi:2013nva,Chung:2023xcv,Ballesteros:2021bee,Caputo:2023ikd,DalCin:2023uai,Takahashi:2019pqf}.

  Our exploration of the complex phenomenology of  axion isocurvature signatures in this light mass range still applies for other sources of axion isocurvature modes. We provide useful analytic scaling relations for the temperature anisotropies across the whole mass range.

\begin{acknowledgments}
R.L. \& W.H. are supported by U.S.\ Dept.\ of Energy contract DE-SC0009924 and the Simons Foundation. D.G. acknowledges support from the Charles Kaufman foundation through grant KA2022-129518, and the provost's office at Haverford College. Portions of the analysis, validation, and figure-preparation code in this work, as well as this acknowledgment, were developed with the assistance of AI coding tools (Claude Code, Anthropic; Codex, OpenAI). Numerical results were independently tested for scientific accuracy  by the authors, who take full responsibility for the scientific content.

\end{acknowledgments}

\appendix

\section{\axie for Isocurvature Modes}
\label{sec:boltz}

In this Appendix we describe the extensions to \axie required to accurately evolve axion isocurvature perturbations and 
compare the results with previous approaches. Although the underlying ETA approach is identical to that used for adiabatic perturbations, isocurvature modes place additional demands due to their reliance on accuracy in the axion dynamics alone, rather than being weighted by the other components of the Universe through metric fluctuations.   We therefore revisit the criteria for the field-to-fluid transition, derive improved initial conditions and validity tests for the ETA approximation, and quantify the accuracy of the resulting implementation. We also document the new user-facing functionality introduced in \axie\,v1.1.

\subsection{ETA}
\label{app:ETA}

\axie implements an effective average over field oscillations on the inverse mass timescale from  Ref.~\cite{Passaglia:2022bcr}.
Here the background field $\phi$ and its perturbation $\delta\phi$ are decomposed into
\begin{align}
\phi &= \varphi_c \cos[m_\ax (t-t_*)] + \varphi_s \sin[m_\ax(t-t_*)], \nonumber \\
\delta\phi &=\delta\varphi_c \cos[ m_\ax(t-t_*)] + \delta\varphi_s \sin[ m_\ax (t-t_*)],
\end{align}
where $t_*$ is the switch epoch that corresponds to a given $H_*/m_\ax$. This redundant system is assigned a unique solution by assuming that the coefficients in the decomposition vary slowly compared with the inverse mass scale, namely 
\begin{equation}
\varphi_c'' = {\cal O}(H/m_\ax) \varphi_c'
\end{equation}
and likewise for the other variables.  Here and below  $'=m_\ax^{-1}d/dt$.
The stress-energy components of the axions, which are quadratic in the field,  are then constructed by analytically averaging $\cos^2 \rightarrow 1/2, \sin^2 \rightarrow 1/2$ and
$\cos \times\sin\rightarrow 0$. After the switch, \axie\ replaces the axion field system with an effective fluid.  Therefore the higher the value of $m_\ax/H_*$, the later the switch and the more accurate the averaging procedure is in general.

While this construction applies to both isocurvature and adiabatic initial conditions when the mass oscillations are the dominant temporal feature, isocurvature initial conditions introduce additional considerations.    Fundamentally, this is because the field fluctuations for isocurvature modes are established at the initial condition rather than being dynamically generated by the curvature fluctuations associated  with the perturbations in other species.   

When only total accuracy of the sum of the adiabatic and isocurvature spectra is required, this usually does not impose additional accuracy considerations.  However to guarantee fractional accuracy of the isocurvature spectra on
their own, we require more stringent control.
We consider accuracy requirements on the switch $m_\ax/H_*$ value in the next section and recall from Eq.~(\ref{eqn:perturbedIC}) that we now always require $m_\ax/H \lesssim 0.1$ at the initial epoch.

Here we address an additional consideration that applies to isocurvature modes for $k\gg k_J$ at $t_*$.  In this regime the Klein-Gordon equation
for perturbations takes the form of a wave equation and in the absence of metric sources is 
\begin{equation}\label{eq:freeosc}
 \delta\phi'' + 3 \frac{H}{m_\ax} \delta\phi'
 + \left( 1+ \frac{k^2}{a^2 m_\ax^2}\right) \delta\phi=0.
\end{equation}
For $k/am_\ax \gg 1$ and $k/aH \gg 1$, this free field equation has oscillatory solutions \begin{equation}
\delta\phi \propto a^{-1} \cos( k\tau+\alpha),\label{eq:free_b}\end{equation} where  $\alpha$ is a constant phase and $\tau$ is the conformal time. For adiabatic fluctuations this equation is sourced by the metric fluctuations and the solution reaches a quasistatic equilibrium form whose timescale is instead dominated by the mass scale oscillations in the background (see Ref. ~\cite{Liu:2024yne} Appendix C), not the faster $k$-scale oscillations.  In the absence of metric sources, these free field oscillations can become much more rapid than the mass scale and invalidate the effective time average approach.   Note that this problem cannot be eliminated simply by increasing $m_\ax/H_*$.  $\delta\varphi_{c,s}$ no longer varies only on the Hubble timescale but instead  $\delta\varphi_{c,s}' = {\cal O}(k/am_\ax) \delta\varphi_{c,s}$, which will always run away for sufficiently large $k$ for any $m_\ax/H_*$ as can be seen from Eq.~(\ref{eq:freeosc}).

Since this consideration is more general than the specific isocurvature case we consider here, in \axie v1.1, we now apply an extra global criterion that checks for the validity of the effective time average.  \axie now only implements this averaging procedure  if
\begin{equation}
\label{eq:noETA}
\frac{(\delta\varphi_c')^2 + (\delta\varphi_s')^2}
{(\delta\varphi_c)^2 + (\delta\varphi_s)^2} < {10}^2,
\end{equation}
where the specific number $10^2$ here has been set at the transition point between the ${\cal O}(1)$ errors induced by not averaging over the phase of oscillations and the growing error due to the fast Jeans oscillations.

  When this check fails, \axie reverts to taking the instantaneous perturbation field values as the stress energy components of the effective fluid perturbations. We find that for adiabatic perturbations the condition never fails regardless of how large $k$ becomes.   For isocurvature perturbations, it fails for $k\gg k_J$ at $t_*$, where the isocurvature perturbations themselves are suppressed and usually entirely negligible compared with adiabatic fluctuations.   When it does, the fractional accuracy of the isocurvature piece depends on the specific phase at which the background field is captured at $t_*$, since the instantaneous density perturbation is the product of the background field and its perturbation.  For our chosen values of $m_\ax/H_*$ this provides a good order-of-magnitude approximation, but we caution the user that special cases when the background $\dot\phi$ instantaneously vanishes can underestimate the already highly suppressed fluid density perturbations.  We show the typical performance in the next section.

\begin{figure}
    \centering
    \includegraphics[width=1\linewidth]{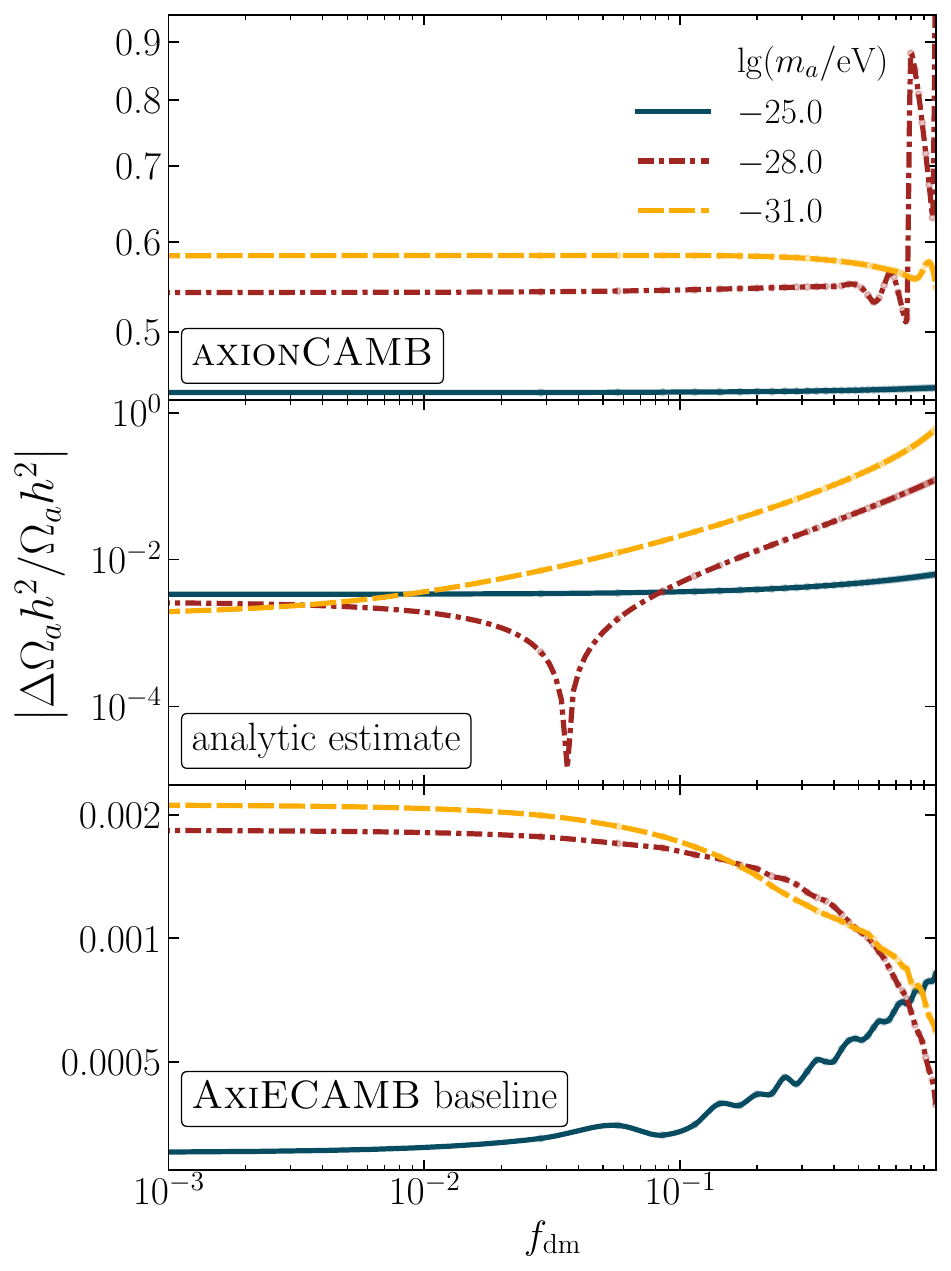}
    \caption{Fractional error in the predicted axion abundance vs our high accuracy results for lower accuracy techniques: \axionCAMB (top), analytic estimates (middle) and \axie baseline (bottom), 
     for a fixed initial field $\phi_\ini$. 
    \axie baseline is already highly accurate and the analytic estimates are accurate at the $10^{-2}$ level for all allowed cases (cf.~Fig.~\ref{fig:rbounds_and_S8}) and better than ${\cal O}(1)$ for all cases.  \axionCAMB is never better than ${\cal O}(1)$ and propagates this error directly to inflationary constraints.  
}
    \label{fig:omaxh2_accuracy_compare}
\end{figure}

\begin{figure}
    \centering
    \includegraphics[width=1\linewidth]{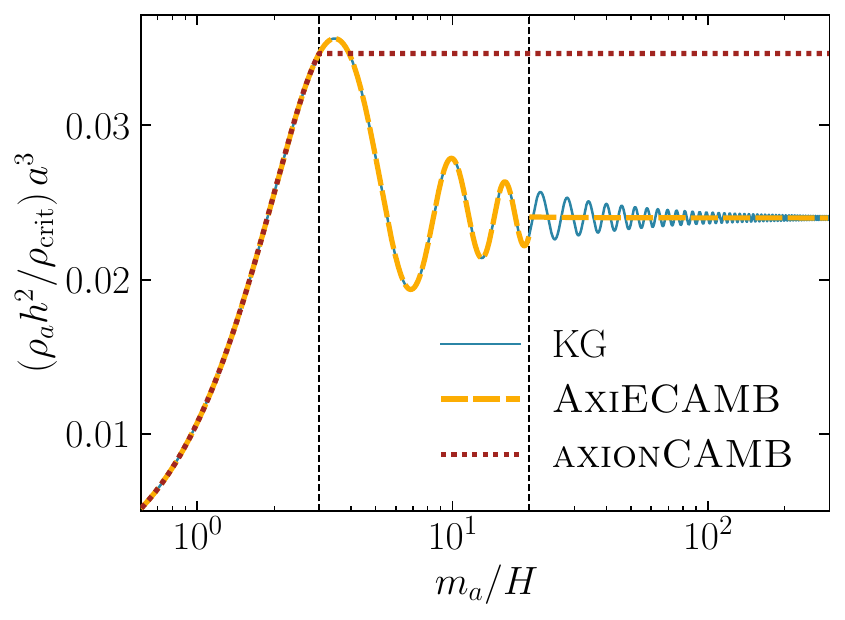}
    \caption{Background axion density evolution $\rho_\ax a^3$ (in units of $\rho_{\rm crit}/h^{2}$),  versus $m_\ax/H$ for $m_\ax = 10^{-25}\,$eV and $f_\dm = 0.2$, a case inside the compatibility window (green shaded) of Fig.~\ref{fig:rbounds_and_S8}. The initial field value $\phi_\ini$ is fixed from the Klein-Gordon reference solution (KG, solid blue, run with \axie at high accuracy).
     \axie (gold dashed) tracks the time average of the KG, while \axionCAMB (red dotted) is off by ${\sim}44\%$, with implications for tensor modes in the compatibility window.  Vertical lines denote their respective switches at $m_\ax/H=20,3$. 
}
    \label{fig:AxiEvsaxC_fixphii}
\end{figure}

\begin{figure}
    \centering
    \includegraphics[width=1\linewidth]{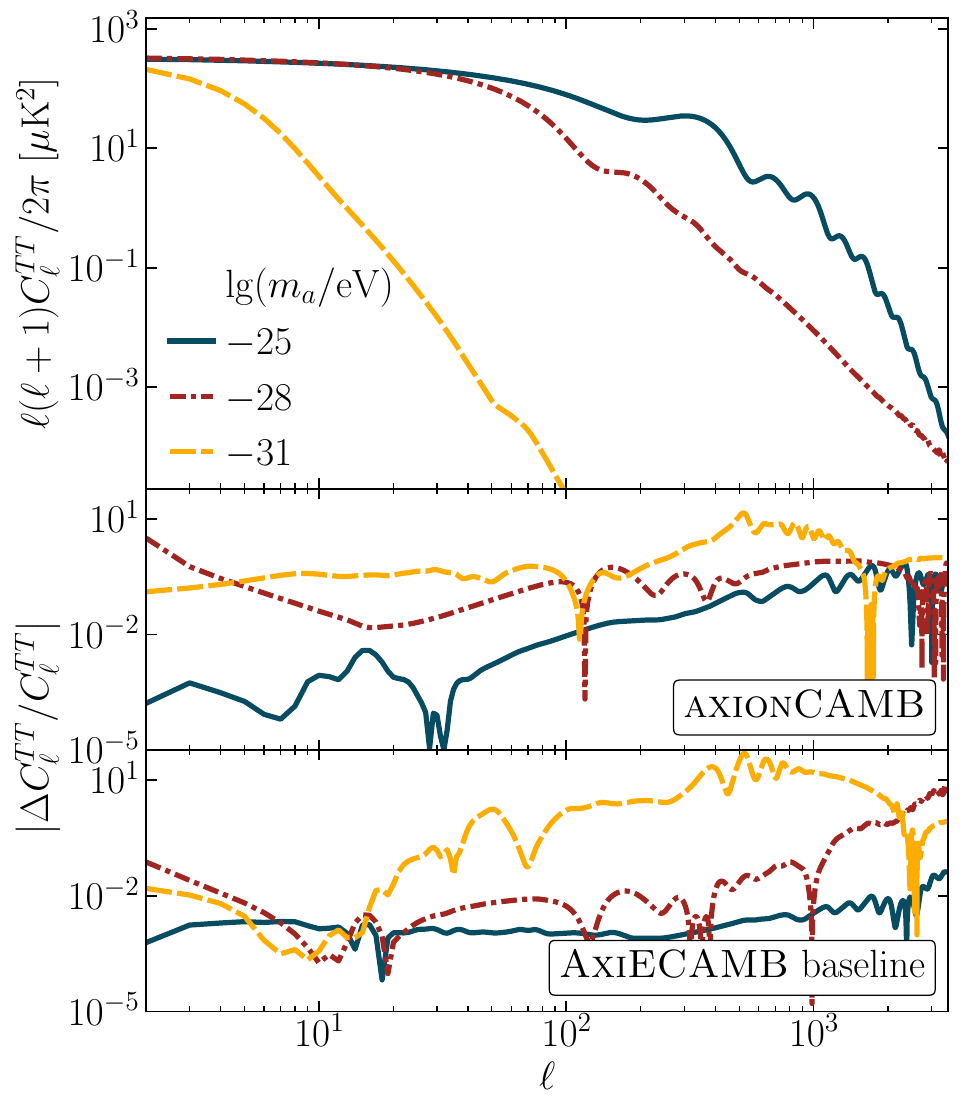}
    
    \caption{Isocurvature $C_\ell^{TT}$ spectrum (top) and its fractional error in lower accuracy techniques (middle \axionCAMB; lower \axie baseline) as in Fig.~\ref{fig:omaxh2_accuracy_compare} for $f_\dm =0.05$ and $A_\iso=1.6\times10^{-7}$.  
    The fractional error of \axie baseline is lower than that of \axionCAMB except for the $\ell\gg \ell_J$ regions where the signal is already highly suppressed.
    }
    \label{fig:ClTT_errors_fdm0d05}
\end{figure}

\begin{figure}[t]
    \centering
    \includegraphics[width=1\linewidth]{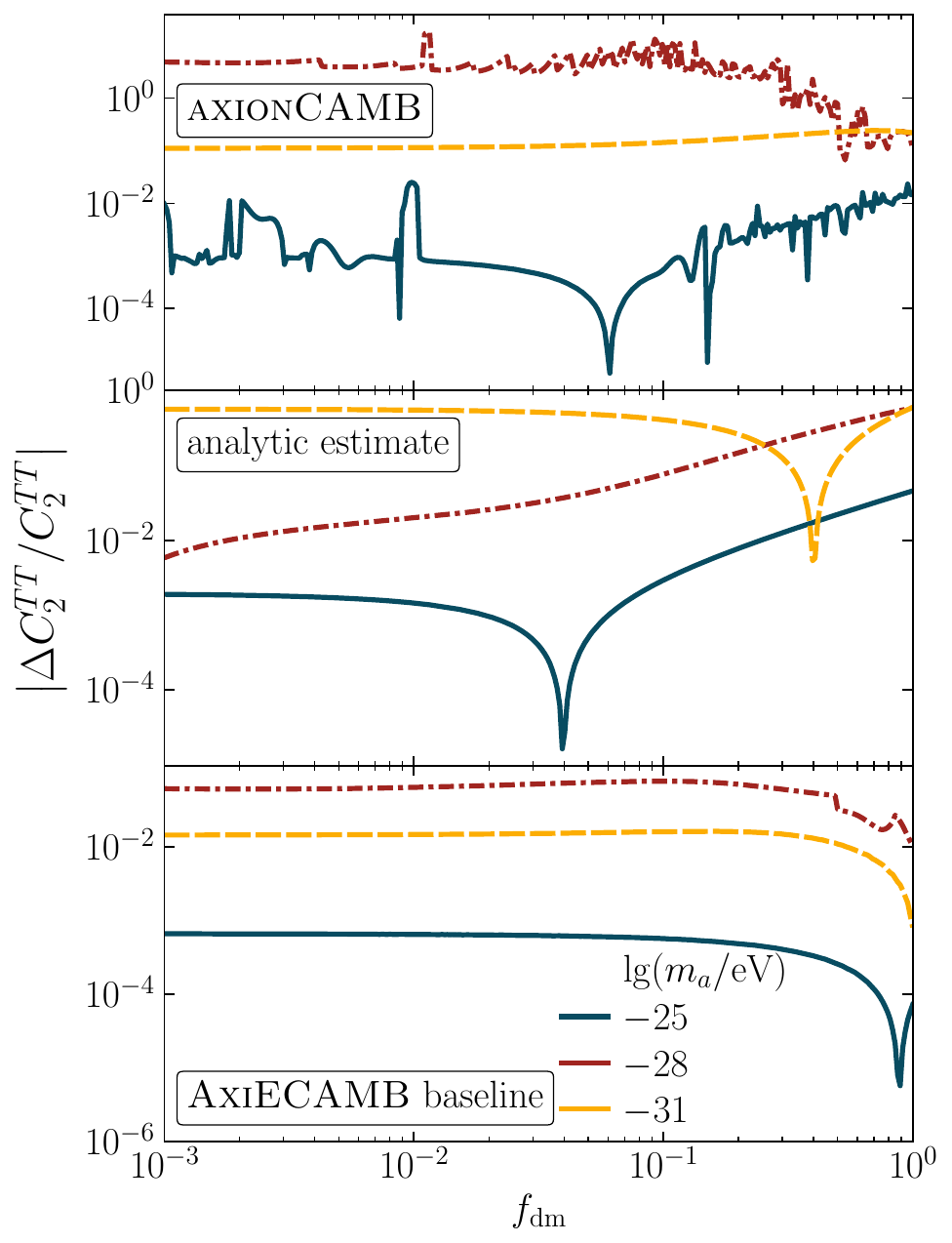}
    \caption{Fractional accuracy of the temperature quadrupole power for various lower accuracy techniques as in Fig.~\ref{fig:omaxh2_accuracy_compare}.  $C_2^{TT}$ provides the most stringent test of accuracy requirements for isocurvature compensation in superhorizon modes and $m_\ax \sim 10^{-28}$\,eV provides the most problematic mass range (see text).   \axie baseline fractional errors are $<0.1$ for any $f_\dm$ even for these cases whereas \axionCAMB errors exceed unity.  The analytic estimate with $C_P = 0.32$ (middle) is better at low $f_\dm$ and higher $m_\ax$ but is better than ${\cal O}(1)$ everywhere.  Since the quadrupole arises from delicate superhorizon compensation, it provides the most stringent accuracy test of the switch algorithm.}
    \label{fig:C2_accuracy_compare}
\end{figure}
\subsection{Accuracy Tests}
\label{app:accuracy}

\subsubsection{$m_\ax>10 H_0$}

For $m_\ax >10 H_0$, the switch from solving the Klein-Gordon equations and outputting instantaneous values of axion quantities to ETAs occurs before or at the present epoch.   After this switch, the axion system is treated as an effective fluid as detailed in Ref.~\cite{Liu:2024yne}.   This switch is controlled by the setting of $m_\ax/H_*$ and if the chosen value is only achieved after the present epoch $H_*<H_0$, the time average is taken at the present epoch and only impacts the definition of the axion quantities and cosmological parameters, not their evolution.

The accuracy of predicted axion observables given the choice of the switch $m_\ax/H_*$ was extensively studied for adiabatic initial conditions in Ref.~\cite{Liu:2024yne} and compared with the previous version of \axionCAMB v2.\footnote{\axionCAMB v2 implemented the switch at $m_\ax/H_*=3h$ rather than the value in its documentation.  \axionCAMB v2.1 corrects this to $m_\ax/H_*=3$ but retains the instantaneous rather than ETA approach.}
Though the basic methodology for the dynamics remains unchanged for isocurvature initial conditions, there are additional accuracy considerations due to the delicate balance between components required to maintain the isocurvature balance of canceling density fluctuations for superhorizon perturbations.   In addition, we test the accuracy of the various scaling relations derived in the main text for the abundance and TT power spectrum.  In summary, in this Appendix we test
\begin{itemize}
\item \axie\,v1.1, $m_\ax/H_*=10$, AccuracyBoost=1 (code release, baseline)
\item \axionCAMB\,v2.1, corrected vs.~v2.0 to switch instantaneously at $m_\ax/H_*=3$
\item Analytic estimates, based on the $\Lambda$CDM expansion history
\end{itemize}
against
\begin{itemize}
\item \axie\,v1.1, $m_\ax/H_*=20$, AccuracyBoost=2 (default for this paper unless otherwise specified)
\end{itemize}

The adiabatic spectra throughout are lensed with \texttt{l\_max\_scalar = 6600}\footnote{This is to distinguish from the $\ell_{\rm max}$ in the Fisher calculation, which is the maximum $\ell$ that the Fisher matrix is summed to in Eq.~(\ref{eq:Fisher}).} and \texttt{k\_eta\_max\_scalar = 18000}, and recombination and  nonlinear CMB lensing are computed respectively with RecFast and the Takahashi halofit \cite{Takahashi:2012em} model. When using \camb v1.6.6, we also adopted RecFast and Takahashi (see \cite{Liu:2025bss} for the impact of reionization and nonlinear lensing assumptions). 

\begin{figure}
    \centering
    \includegraphics[width=1\linewidth]{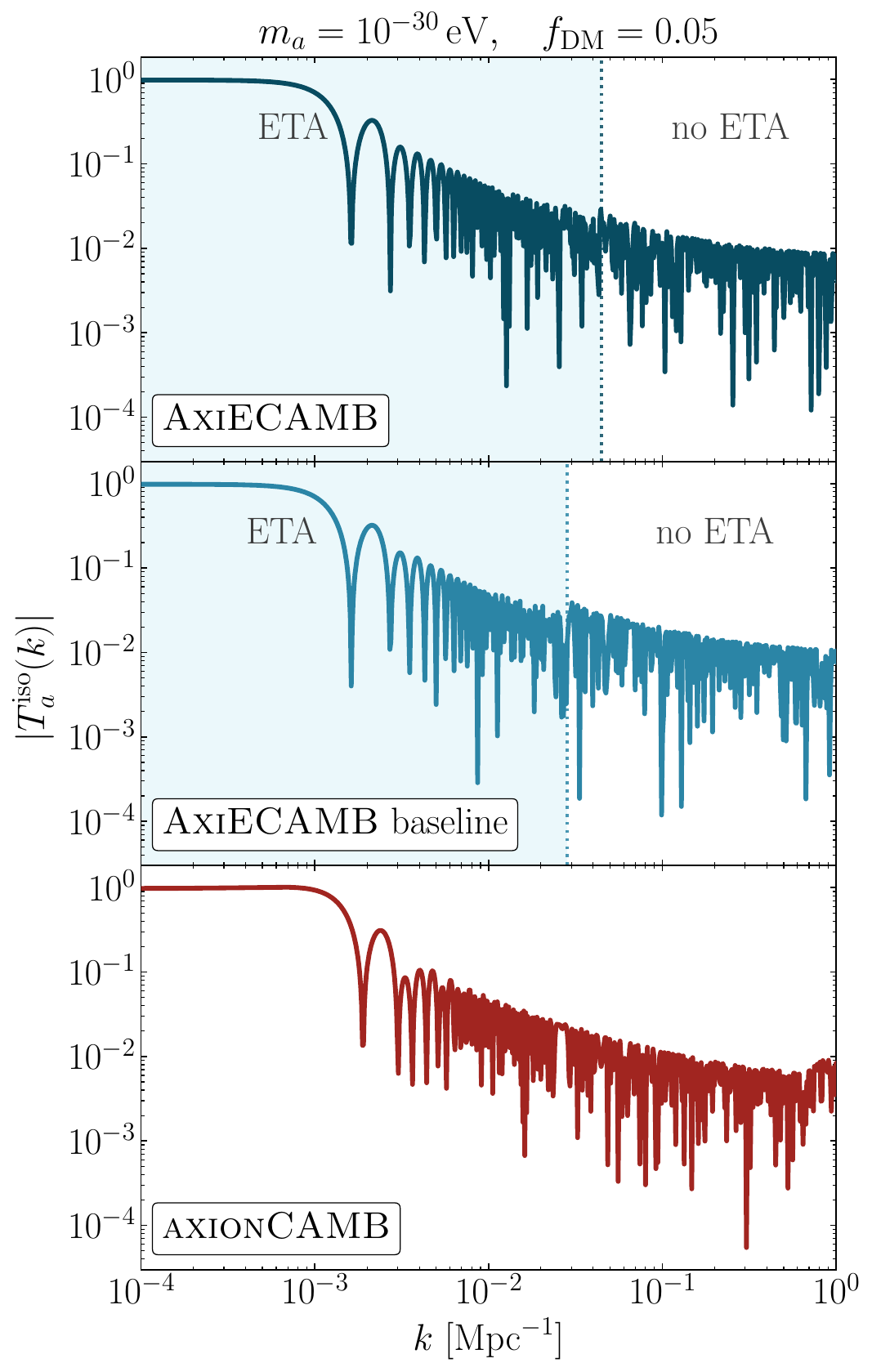}
    \caption{Axion isocurvature transfer function $|T_\ax^{\iso}(k)|$ with the various codes and accuracy settings as in Fig.~\ref{fig:ClTT_errors_fdm0d05}.  Here $m_\ax=10^{-30}\,{\rm eV}$ and
$f_\dm=0.05$. Top to bottom: \axie, \axie\ baseline, \axionCAMB. The dotted lines indicate the wavenumber at which \axie\ turns off ETA due to  Jeans oscillations exceeding the mass-oscillation in frequency (see Eq.~(\ref{eq:noETA})).  In the highly suppressed no-ETA range, \axie\ traces the oscillation envelope instead of its average, and is not as accurate. \axionCAMB\ has no ETA which makes it less accurate in the ETA region but also not subject to a transitional glitch to no ETA. 
}
    \label{fig:noETA_demo}
\end{figure}

\begin{figure}
    \centering
    \includegraphics[width=1\linewidth]{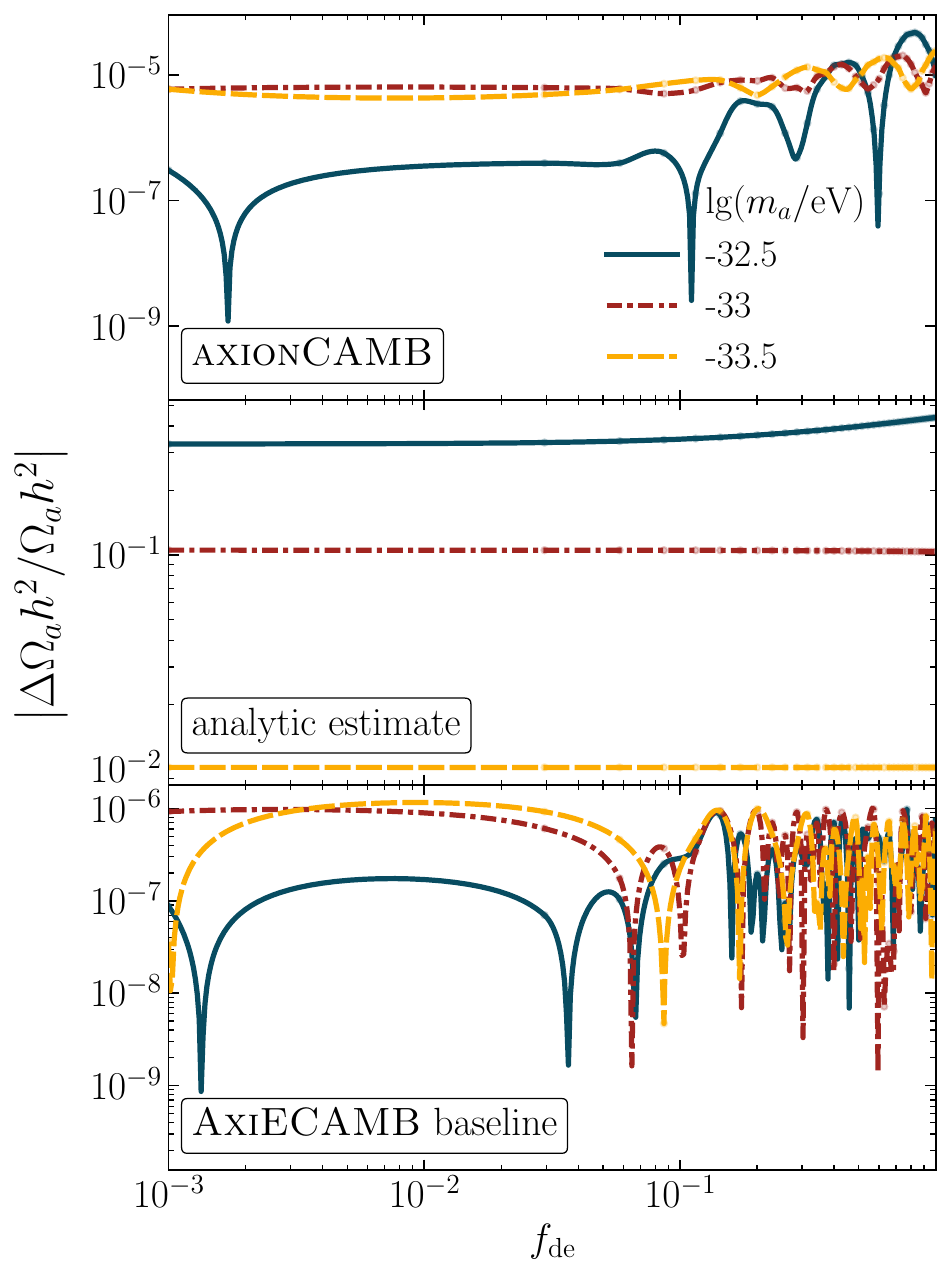}
    \caption{Fractional error in the predicted axion abundance as in Fig.~\ref{fig:omaxh2_accuracy_compare} but for dark-energy masses and fractions.  Both \axie baseline and \axionCAMB are accurate at the ${\cal O}(10^{-5})$ level.   Analytic estimate errors increase for the higher masses since \axie outputs instantaneous $\Omega_\ax h^2$ instead of the time-average for $m_\ax<10 H_0$.}
    \label{fig:omaxh2_accuracy_compare_DE}
\end{figure}

 We begin with tests of the axion abundance given the initial field $\phi_\ini$ in Fig.~\ref{fig:omaxh2_accuracy_compare}. They are especially relevant for isocurvature modes and their constraints on inflation since the mapping between $A_\iso$ and $r_\ax$ involves the initial field through Eq.~(\ref{eq:AtoR}).
Here the \axie baseline is already accurate at the $10^{-3}$ level or better for all $m_\ax$~and $f_\dm$.  Our analytic estimate from Eq.~(\ref{eq:F}) for the abundance is accurate at better than the $10^{-2}$ level for the high mass $m_\ax\gg H_\eq$ case for all $f_\dm$ and for the lighter masses that accuracy level for all allowed $f_\dm \lesssim 0.05$. 
 Even for the highly ruled out case of $f_\dm=1$ the error never exceeds unity and remains a good scaling estimate for the abundance.

In contrast, for \axionCAMB, the error in $\Omega_\ax h^{2}$ is ${\cal O}(1)$ for all $m_\ax$ and $f_\dm$ for all species. This behavior is expected for its instantaneous switch to the EFA after $m_\ax/H=3$. 
In Fig.~\ref{fig:AxiEvsaxC_fixphii} we illustrate the origin of this error by showing the time evolution of $\rho_\ax a^3$ from the codes.   Notice that $m_\ax/H=3$ catches the first oscillation of the axion field near its extrema and therefore produces a large error in  $\rho_\ax$ afterwards.   \axie not only has a later switch value when the oscillations are smaller but also takes the ETA of the oscillations and matches the Klein-Gordon solution to high accuracy.
For \axionCAMB, the anomalous behavior for $
m_\ax=10^{-28}$\,eV and $f_\dm \sim 1$ is due to a bug in the implementation of spline interpolation which appears at this $m_\ax$ because of the range chosen for the spline table.

We next examine the fractional error $\Delta C_\ell^{TT}/C_\ell^{TT}$ for these $m_\ax$ and a fixed $A_\iso$ in Fig.~\ref{fig:ClTT_errors_fdm0d05}. Fixing $A_\iso$ decouples the error in abundance $\rho_\ax$ at fixed
$\phi_\ini$ from the errors due to perturbation evolution with a switch. Because $C_\ell^{TT}$ scales linearly with $A_\iso$ in each code, the fractional errors $\Delta C_\ell^{TT}/C_\ell^{TT}$ shown here are independent of the chosen $A_\iso$; here we use the fiducial $A_\iso=1.6\times10^{-7}$ to match the related figures in the main text. Baseline \axie precision settings yield accurate isocurvature predictions for intermediate $\ell$ before the Jeans suppression, especially where cosmic variance is small. The fractional errors $\Delta C_\ell^{TT}/C_\ell^{TT}$ can be large for much larger multipoles where the signal is highly suppressed.  This is in part due to the baseline setting of AccuracyBoost=1, especially for $m_\ax\ll H_\eq$. For isocurvature work in this paper, we have therefore set AccuracyBoost=2.

The rise in the fractional error toward the lowest $\ell$ values with the \axie baseline is notable for $m_\ax <H_\eq$, although it is still well below  cosmic variance. Indeed,  $C_2^{TT}$ at $m_\ax \sim H_\eq \sim 10^{-28}$\,eV typically is the most stringent test of code accuracy for both \axie and \axionCAMB, imposing the tightest requirement for low-$k$ perturbation evolution, as the modes are super-horizon, with metric fluctuation evolution at the switch directly impacting observables.  For much higher $m_\ax$, the switch precedes the epoch where perturbations   impact CMB observables, while for lower $m_\ax$ the horizon grows such that $k/aH$ of a given $k$-mode is larger.

Superhorizon axion isocurvature fluctuations require high accuracy as they must very precisely cancel fluctuations in non-axion species.
This compensation balance is essentially the consequence of a separate universe mechanism, as the evolution of superhorizon modes is indistinguishable from the local background evolution. A higher local
$H$ induces more redshift in other components thus yielding a compensating underdensity to maintain the initial isocurvature condition.  Curvature fluctuations are only generated as $(k/aH)^2\delta_\ax(k,\tau_\ini)$ near horizon crossing. While the switch to the ETA maintains this balance to order $(H/m_\ax)^2 \delta_\ax(k,\tau_\ini)$, the required $(k/aH)^2$ factor in the compensation means that the fractional error from the switch will run away as $k\rightarrow 0$. This makes the largest observationally relevant modes the strongest test of the switch algorithm for isocurvature modes.
We thus set a higher $m_\ax/H_*=20$ for the switch than in  baseline \axie.

\begin{figure}
    \centering
    \includegraphics[width=1\linewidth]{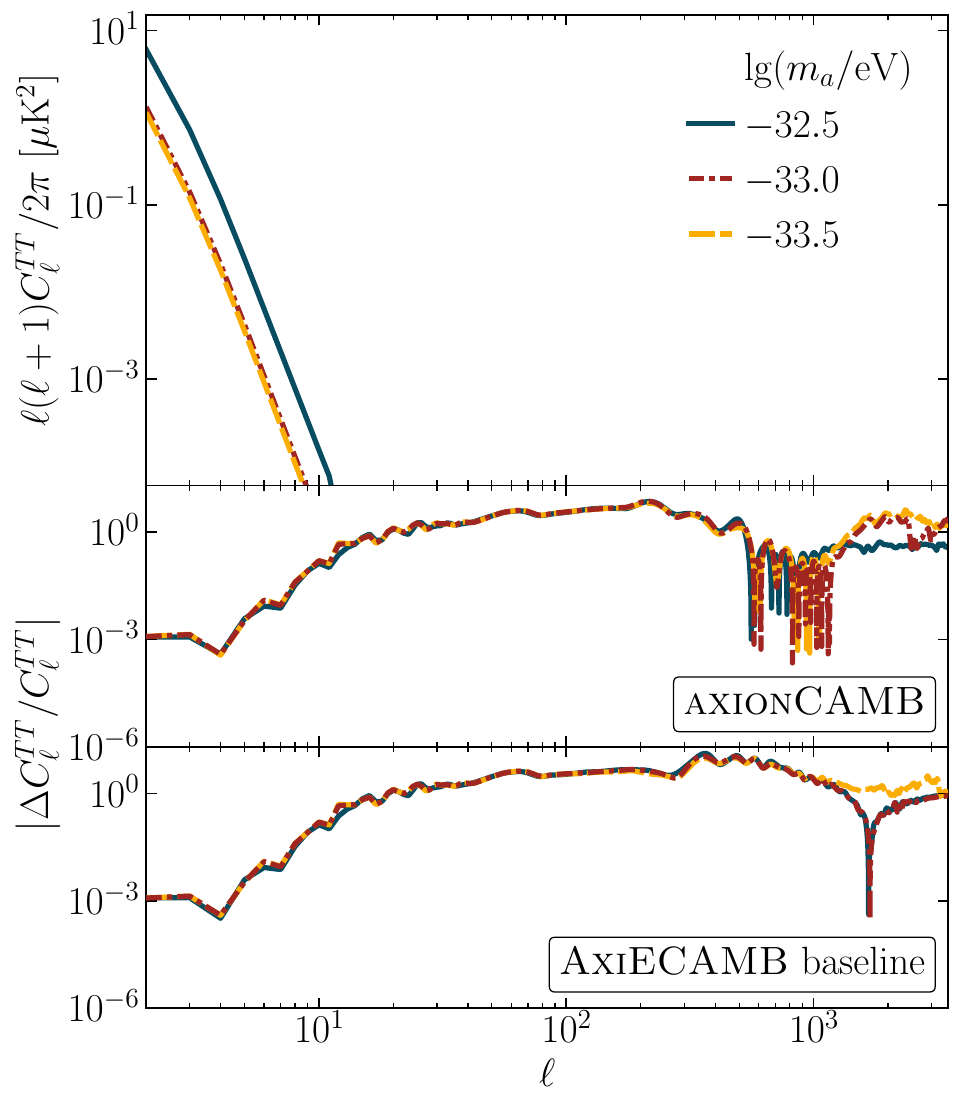}
    \caption{Isocurvature $C_\ell^{TT}$ spectrum (top) and its fractional error in lower accuracy techniques (middle \axionCAMB; lower \axie baseline) as in Fig.~\ref{fig:ClTT_errors_fdm0d05} but for dark-energy masses and $A_\iso=1.6\times10^{-7}$, $f_\de=0.1$. Both \axie baseline and \axionCAMB are sufficiently accurate for the observationally relevant low-$\ell$ regime but fractional errors increase in the highly suppressed regime.}
    \label{fig:ClTT_errors_fde0d1}
\end{figure}

\begin{figure}
    \centering
    \includegraphics[width=1\linewidth]{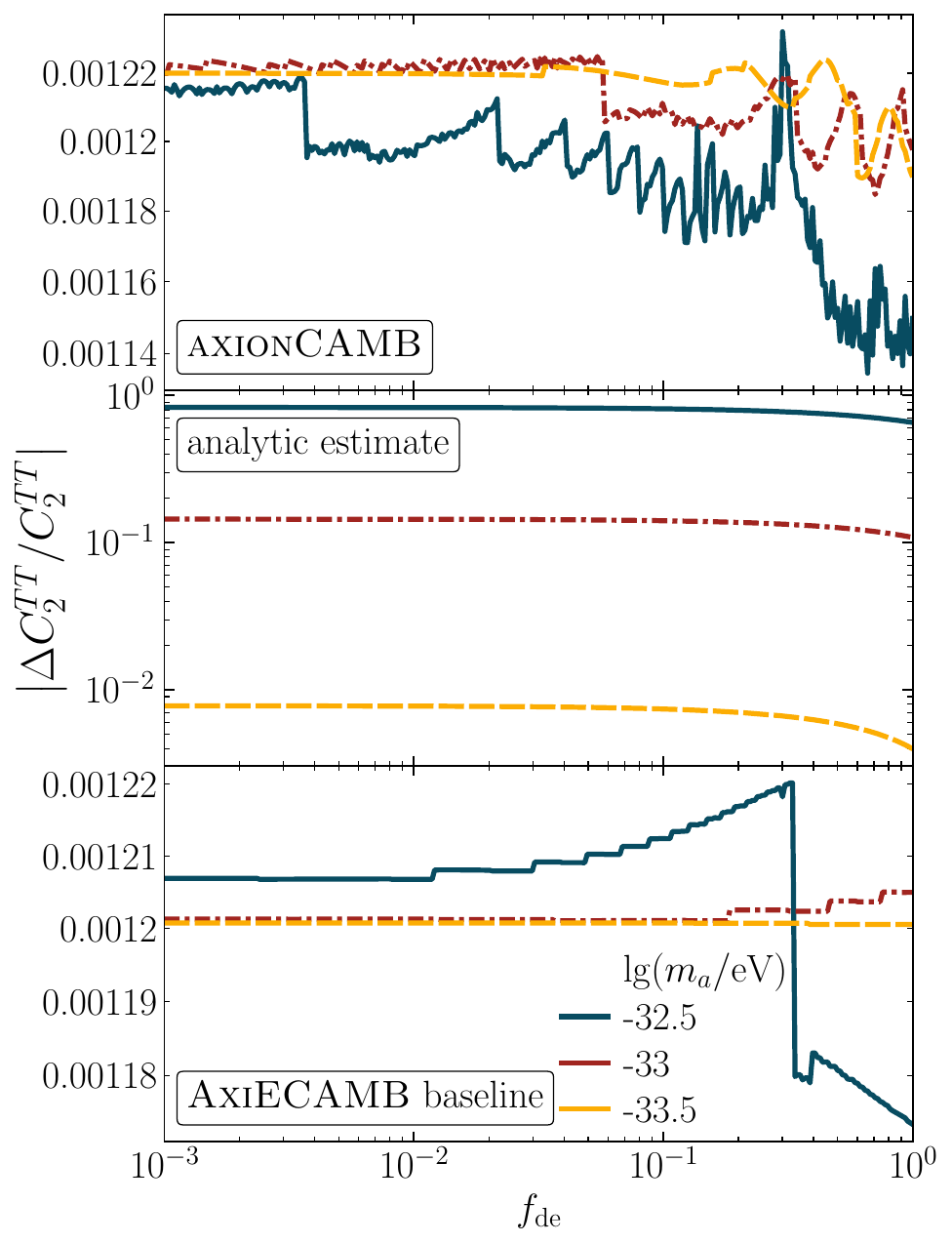}
    \caption{Fractional accuracy of the temperature quadrupole as in Fig.~\ref{fig:C2_accuracy_compare} but for 
    dark energy masses and fraction $f_\de$.  Both \axie baseline and \axionCAMB are highly accurate.  For the analytic estimate with $C_Q = 0.0097$, the accuracy is high for the lowest masses but degrades toward the edge of the dark energy regime where the background field is no longer frozen.
}
    \label{fig:C2_accuracy_compare_DE}
\end{figure}

In contrast, \axionCAMB is only $\sim 10\%$ accurate at observationally relevant $\ell$, though it can outperform baseline accuracy when $\ell \gg \ell_{J}$.  
Although \axionCAMB does not introduce a direct violation of isocurvature density compensation due to the instantaneous and continuous (in density) switch, there is a pressure discontinuity introduced by using $p_\ax=0$ after $m_\ax/H=3$ (see e.g. \cite{Liu:2024yne} Eq. ~(22)) which has similar consequences. 

Beyond the fixed value of $f_\dm=0.05$ case in Fig.~\ref{fig:ClTT_errors_fdm0d05}, we show the accuracy of this stringent test at $\ell=2$ as a function of $f_\dm$ in 
Fig.~\ref{fig:C2_accuracy_compare}.  Even for the worst case, the baseline \axie accuracy for $m_\ax \sim 10^{-28}$\,eV is ${{\cal O}}(10^{-1})$ even for $f_\dm\rightarrow 0$ whereas it is better for both the heavier and lighter $m_\ax$ cases.   In contrast \axionCAMB errors substantially exceed ${\cal O}(1)$.

In addition, we test the analytic estimate of $C_2^{TT}$ from $C_P$ from Eq.~(\ref{eqn:CP_for_plateau}), derived under the assumption that the expansion history remains close to that of $\Lambda$CDM.  For the $m_\ax=10^{-25}~{\rm eV}$ and $10^{-28}~{\rm eV}$ cases the estimate is accurate at the fractional ${\cal O}(10^{-2})$ level in the allowed range of $f_\dm$. Even for the excluded region $m_\ax=10^{-28}~{\rm eV}, f_\dm \rightarrow 1$, the estimation errors never exceed ${\cal O}(1)$.
For $10^{-31}$\,eV and below, the Jeans suppression starts to affect the quadrupole which is not accounted for in the analytic formula.

Finally, we test the accuracy of the isocurvature axion transfer function in Fig.~\ref{fig:noETA_demo}.  As discussed in the previous section we perform the effective time average over mass timescale oscillations only if Eq.~(\ref{eq:noETA}) is satisfied and the field evolves predominantly on that scale. The consequence is that the isocurvature calculation is less accurate at $k\gg k_J$, due to the oscillatory changes in the field background modulating the density perturbations. There the axion density perturbations are already small and do not affect the $C_\ell$ over the $\ell$ range used in this work. This lower accuracy, while good enough for parameter estimation with adiabatic modes, can itself be mitigated by raising $m_\ax/H_*$ to push the transition to higher $k$, as shown in the comparison to the baseline case.  
We also show the \axionCAMB results where the switch is before the oscillations begin in the background, and hence do not have the same problem at $k\gg k_J$, but correspondingly sacrifice accuracy in the more important $k\lesssim k_J$ region.

These techniques and compensation considerations also apply when the Peccei-Quinn symmetry is broken after the end of inflation, though isocurvature fluctuations take on a superhorizon white-noise power spectrum in that case. In this case, \axie may be used to compute observables so long as the isocurvature perturbations are on length scales much larger than the maximal free streaming scale for field perturbations \cite{Liu:2025lts,Amin:2025sla} and the axion background density and  the desired white-noise power spectrum are appropriately matched (see e.g.~\cite{Harigaya:2025pox}).

\subsubsection{$m_\ax\le 10 H_0$}

For $m_\ax\le 10 H_0$, \axie  integrates the exact Klein-Gordon equations to the present and also uses the instantaneous (rather than ETA) abundance to define cosmological parameters. We test the accuracy of the baseline \axie settings, analytic estimates for the abundance and quadrupole as well as \axionCAMB against the default \axie computations.

For the abundance, the lower accuracy settings of \axie baseline lead to a negligible effect, as expected.  For the chosen masses, \axionCAMB also solves the Klein-Gordon equation all the way to the present and is highly accurate. 

For $m_\ax \sim 10^{-33}~{\rm eV}$ and $10^{-32.5}~{\rm eV}$, the analytic estimate from Eq.~(\ref{eq:F}) differs at the $10^{-1}$ level for any $f_\de=\Omega_\ax/\Omega_\de$ (see Fig.~\ref{fig:omaxh2_accuracy_compare_DE}). This occurs because the analytic estimate is meant to track the time-averaged abundance rather than the instantaneous value (see Fig.~\ref{fig:F_and_fit}). Since the abundance itself scales with $f_\de$ across the wide range of its values, the analytic formula remains an excellent scaling relation in $f_\de$ even for these masses.

In Fig.~\ref{fig:ClTT_errors_fde0d1}, we show the accuracy of $C_\ell^{TT}$ for the various codes.  Since both \axie baseline and \axionCAMB solve the KG equations directly to the present for the $m_\ax$ values shown, observables are highly accurate for the lowest $\ell$ and suffice to estimate parameters for adiabatic-isocurvature mixtures. Due to suppression beyond the Jeans scale, both codes show rising $\Delta C_\ell^{TT}/C_\ell^{TT}$ isocurvature errors at higher multipoles that would be masked by adiabatic contributions.

The analytic estimate from $C_Q$ is also highly precise for $m_\ax<10^{-33.5}~{\rm eV}$ as it is derived for a frozen background field, which we show in Fig.~\ref{fig:C2_accuracy_compare_DE}.
For larger $m_\ax$, the estimate is inaccurate at the $10^{-1}$ to ${\cal O}(1)$ level due to the background and perturbations beginning to evolve before the present. The perturbation errors result from the interpolation between the $C_Q$ and $C_P$, $\ell=2$ values, as seen for the $m_\ax= 10^{-31}~{\rm eV}$ case and the $m_\ax$ values in Fig.~\ref{fig:fDM0d02_fixedbetaiso}. The  $C_Q$ estimate matches the scaling of the $f_\de^2$ dependence of $\ell=2$ observables  for all $m_\ax\lesssim 10^{-32.5}~{\rm eV}$ whereas the $C_P$ estimate matches the $f_\dm^2$ scaling for all $m_\ax \gtrsim 10^{-31}~{\rm eV}$.

\subsection{Isocurvature Parameters and Interface}

\axie\ computes axion isocurvature spectra from its separate initial condition, normalized through the axion
transfer function of Eq.~(\ref{eq:T_i_iso}), activated by
\texttt{axion\_isocurvature = T}. Unlike in \axionCAMB, adiabatic and isocurvature CMB power spectra are
computed in separate runs and should be summed externally, which is exact for the uncorrelated
isocurvature scenario where no cross spectra arise. 
By default, we disable the fiducial-template $C_\ell$ interpolation,
\texttt{use\_spline\_template = F}, for both adiabatic and isocurvature runs, consistent with isocurvature spectra never being interpolated against an adiabatic template.

We expand the user-input parameters in \texttt{params.ini} so that the primordial isocurvature spectrum
$\Delta^2_\iso(k) = A_\iso (k/k_0)^{n_\ax}$ with $k_0 = 0.05~{\rm Mpc}^{-1}$
may be specified in three ways, selected by an integer \texttt{iso\_input\_mode}:
\begin{enumerate}\setlength\itemsep{3pt}\setlength\parskip{6pt}
    \item $(A_\iso,\, n_\ax)$ via \texttt{A\_iso}, \texttt{n\_a\_iso}: direct
phenomenological input of the amplitude at $k_0$, independent of the
inflationary origin (as in
Figs.~\ref{fig:fDM0d02_fixedbetaiso}, \ref{fig:Tiso}).
\item $(r_\ax,\, n_\ax)$ via \texttt{r\_a\_iso}, \texttt{n\_a\_iso}
(default): following Eq.~(\ref{eq:AtoR}),
where $A_\iso$ is derived from $r_\ax$, $A_s$ and the initial field value
$\phi_\ini$. 
\item $(\Hinf,\, n_\ax)$ via \texttt{Hinf} [$\log_{10}(\Hinf/\text{GeV})$]: \axie derives $r_\ax = 2\Hinf^2/(\pi^2 A_s \Mpl^2)$ for a
massless spectator field and then proceeds as in mode 2.
\end{enumerate}

Each mode takes exactly one amplitude input: \texttt{A\_iso}, \texttt{r\_a\_iso} and
\texttt{Hinf} respectively, and ignores the other two. \axie\ relates them through the background solution of $\phi_\ini$, and
\texttt{iso\_consistency = T} imposes the canonical single-field consistency relation
$n_\ax = -r_\ax/8$ for any amplitude mode; if \texttt{F}, $n_\ax$ is taken from \texttt{n\_a\_iso}.
At \texttt{feedback\_level} $\geq 1$, \axie now reports the derived initial field value in Planck units, $\phi_\ini/\Mpl$, together with the isocurvature amplitude expressed in all three parameterizations $A_\iso$, $r_\ax$ and $\Hinf$.

Beyond the isocurvature extension, we introduce additional improvements to \axie for adiabatic modes and the background. We make the KG-to-EFA switch point, i.e. $m_\ax/H_*$, a user-input parameter, \texttt{movH\_switch}, with baseline value 10 (smaller values are not supported). Larger values track more exact KG oscillation cycles and improve accuracy only when paired with a corresponding increase in \texttt{accuracy\_boost} so that the rapid oscillations remain resolved in time. While the settings used in this work (\texttt{movH\_switch =
20}, \texttt{accuracy\_boost = 2}) are validated in
Appendix~\ref{app:accuracy}, for other choices, the user should increase \texttt{accuracy\_boost} at
fixed \texttt{movH\_switch} until their target observables are stable, and confirm insensitivity to a further increase of \texttt{movH\_switch}. 

Nonlinear CMB lensing based on Halofit is available from \axie\,v1.1 onward.  Previously it was turned off due to the problem of spline interpolating a sparsely sampled  matter transfer function in the highly suppressed Jeans oscillation regime when axions dominate the dark matter.   In v1.1, this highly suppressed regime is treated under linear theory for lensing. We caution that Halofit itself has not been recalibrated for axions.  For the results here the impact is negligible: nonlinear lensing changes the Fisher information for $r_\ax$ ($\ell \leq 2000$) by at most $1.3\%$.
For isocurvature modes we only provide the unlensed spectra, 
consistent with how \camb treats tensor modes, and throughout, the total power spectra refer to the sum of the lensed adiabatic and unlensed isocurvature $C_\ell$. This is because the sum of the adiabatic and isocurvature unlensed power spectra should actually be lensed by the sum of their respective lens power spectra.  For observationally allowed isocurvature fractions, this would be indistinguishable from simply lensing the adiabatic modes.

Throughout this work and from \axie v1.1 onwards, we also fix the convention for the reported axion abundance at
\begin{equation}\nonumber
m_{\rm ETA}\equiv 10H_0^{\rm fid}
=1.44\times10^{-32}\,{\rm eV},
\end{equation}
where $H_0^{\rm fid}$ is the fiducial value adopted in this work, rather than recomputed from
$m_\ax/H_0=10$ for each sampled cosmology. For
$m_\ax < m_{\rm ETA}$, \axie reports the instantaneous present-day $\Omega_\ax h^2$, whereas for $m_\ax\geq m_{\rm ETA}$ it reports the effective time average. Fixing
$m_{\rm ETA}$ in physical units, rather than reevaluating $10H_0$ at each sample, prevents the parameter definition itself from changing when $H_0$ is varied in parameter estimation. This abundance convention is separate from the dynamical KG-to-EFA switch $m_\ax/H_*$ (\texttt{movH\_switch}). 

Finally, \axie outputs the isocurvature transfer function in our convention (\ref{eq:T_i_iso}), which differs from the curvature based \camb convention by $k^2$. The isocurvature tilt \texttt{n\_a\_iso} likewise follows the tensor convention: the scale-invariant $n_\ax=0$ corresponds to $n=1$ in the adiabatic convention for the scalar power spectrum in \camb.
\vfill
\bibliography{main}

\begin{thebibliography}{62}%
\makeatletter
\providecommand \@ifxundefined [1]{%
 \@ifx{#1\undefined}
}%
\providecommand \@ifnum [1]{%
 \ifnum #1\expandafter \@firstoftwo
 \else \expandafter \@secondoftwo
 \fi
}%
\providecommand \@ifx [1]{%
 \ifx #1\expandafter \@firstoftwo
 \else \expandafter \@secondoftwo
 \fi
}%
\providecommand \natexlab [1]{#1}%
\providecommand \enquote  [1]{``#1''}%
\providecommand \bibnamefont  [1]{#1}%
\providecommand \bibfnamefont [1]{#1}%
\providecommand \citenamefont [1]{#1}%
\providecommand \href@noop [0]{\@secondoftwo}%
\providecommand \href [0]{\begingroup \@sanitize@url \@href}%
\providecommand \@href[1]{\@@startlink{#1}\@@href}%
\providecommand \@@href[1]{\endgroup#1\@@endlink}%
\providecommand \@sanitize@url [0]{\catcode `\\12\catcode `\$12\catcode `\&12\catcode `\#12\catcode `\^12\catcode `\_12\catcode `\%12\relax}%
\providecommand \@@startlink[1]{}%
\providecommand \@@endlink[0]{}%
\providecommand \url  [0]{\begingroup\@sanitize@url \@url }%
\providecommand \@url [1]{\endgroup\@href {#1}{\urlprefix }}%
\providecommand \urlprefix  [0]{URL }%
\providecommand \Eprint [0]{\href }%
\providecommand \doibase [0]{https://doi.org/}%
\providecommand \selectlanguage [0]{\@gobble}%
\providecommand \bibinfo  [0]{\@secondoftwo}%
\providecommand \bibfield  [0]{\@secondoftwo}%
\providecommand \translation [1]{[#1]}%
\providecommand \BibitemOpen [0]{}%
\providecommand \bibitemStop [0]{}%
\providecommand \bibitemNoStop [0]{.\EOS\space}%
\providecommand \EOS [0]{\spacefactor3000\relax}%
\providecommand \BibitemShut  [1]{\csname bibitem#1\endcsname}%
\let\auto@bib@innerbib\@empty
\bibitem [{\citenamefont {Svrcek}\ and\ \citenamefont {Witten}(2006)}]{Svrcek:2006yi}%
  \BibitemOpen
  \bibfield  {author} {\bibinfo {author} {\bibfnamefont {P.}~\bibnamefont {Svrcek}}\ and\ \bibinfo {author} {\bibfnamefont {E.}~\bibnamefont {Witten}},\ }\href {https://doi.org/10.1088/1126-6708/2006/06/051} {\bibfield  {journal} {\bibinfo  {journal} {JHEP}\ }\textbf {\bibinfo {volume} {06}},\ \bibinfo {pages} {051}},\ \Eprint {https://arxiv.org/abs/hep-th/0605206} {arXiv:hep-th/0605206} \BibitemShut {NoStop}%
\bibitem [{\citenamefont {Axenides}\ \emph {et~al.}(1983)\citenamefont {Axenides}, \citenamefont {Brandenberger},\ and\ \citenamefont {Turner}}]{Axenides:1983hj}%
  \BibitemOpen
  \bibfield  {author} {\bibinfo {author} {\bibfnamefont {M.}~\bibnamefont {Axenides}}, \bibinfo {author} {\bibfnamefont {R.~H.}\ \bibnamefont {Brandenberger}},\ and\ \bibinfo {author} {\bibfnamefont {M.~S.}\ \bibnamefont {Turner}},\ }\href {https://doi.org/10.1016/0370-2693(83)90586-5} {\bibfield  {journal} {\bibinfo  {journal} {Phys. Lett. B}\ }\textbf {\bibinfo {volume} {126}},\ \bibinfo {pages} {178} (\bibinfo {year} {1983})}\BibitemShut {NoStop}%
\bibitem [{\citenamefont {Turner}\ \emph {et~al.}(1983)\citenamefont {Turner}, \citenamefont {Wilczek},\ and\ \citenamefont {Zee}}]{Turner:1983sj}%
  \BibitemOpen
  \bibfield  {author} {\bibinfo {author} {\bibfnamefont {M.~S.}\ \bibnamefont {Turner}}, \bibinfo {author} {\bibfnamefont {F.}~\bibnamefont {Wilczek}},\ and\ \bibinfo {author} {\bibfnamefont {A.}~\bibnamefont {Zee}},\ }\href {https://doi.org/10.1016/0370-2693(83)91229-7} {\bibfield  {journal} {\bibinfo  {journal} {Phys. Lett. B}\ }\textbf {\bibinfo {volume} {125}},\ \bibinfo {pages} {35} (\bibinfo {year} {1983})},\ \bibinfo {note} {[Erratum: Phys.Lett.B 125, 519 (1983)]}\BibitemShut {NoStop}%
\bibitem [{\citenamefont {Seckel}\ and\ \citenamefont {Turner}(1985)}]{Seckel:1985tj}%
  \BibitemOpen
  \bibfield  {author} {\bibinfo {author} {\bibfnamefont {D.}~\bibnamefont {Seckel}}\ and\ \bibinfo {author} {\bibfnamefont {M.~S.}\ \bibnamefont {Turner}},\ }\href {https://doi.org/10.1103/PhysRevD.32.3178} {\bibfield  {journal} {\bibinfo  {journal} {Phys. Rev. D}\ }\textbf {\bibinfo {volume} {32}},\ \bibinfo {pages} {3178} (\bibinfo {year} {1985})}\BibitemShut {NoStop}%
\bibitem [{\citenamefont {Lyth}(1992)}]{Lyth:1991ub}%
  \BibitemOpen
  \bibfield  {author} {\bibinfo {author} {\bibfnamefont {D.~H.}\ \bibnamefont {Lyth}},\ }\href {https://doi.org/10.1103/PhysRevD.45.3394} {\bibfield  {journal} {\bibinfo  {journal} {Phys. Rev. D}\ }\textbf {\bibinfo {volume} {45}},\ \bibinfo {pages} {3394} (\bibinfo {year} {1992})}\BibitemShut {NoStop}%
\bibitem [{\citenamefont {Fox}\ \emph {et~al.}(2004)\citenamefont {Fox}, \citenamefont {Pierce},\ and\ \citenamefont {Thomas}}]{Fox:2004kb}%
  \BibitemOpen
  \bibfield  {author} {\bibinfo {author} {\bibfnamefont {P.}~\bibnamefont {Fox}}, \bibinfo {author} {\bibfnamefont {A.}~\bibnamefont {Pierce}},\ and\ \bibinfo {author} {\bibfnamefont {S.~D.}\ \bibnamefont {Thomas}},\ }\href@noop {} {\bibfield  {journal} {\bibinfo  {journal} {SLAC-PUB-10030}\ } (\bibinfo {year} {2004})},\ \Eprint {https://arxiv.org/abs/hep-th/0409059} {arXiv:hep-th/0409059} \BibitemShut {NoStop}%
\bibitem [{\citenamefont {Hertzberg}\ \emph {et~al.}(2008)\citenamefont {Hertzberg}, \citenamefont {Tegmark},\ and\ \citenamefont {Wilczek}}]{Hertzberg:2008wr}%
  \BibitemOpen
  \bibfield  {author} {\bibinfo {author} {\bibfnamefont {M.~P.}\ \bibnamefont {Hertzberg}}, \bibinfo {author} {\bibfnamefont {M.}~\bibnamefont {Tegmark}},\ and\ \bibinfo {author} {\bibfnamefont {F.}~\bibnamefont {Wilczek}},\ }\href {https://doi.org/10.1103/PhysRevD.78.083507} {\bibfield  {journal} {\bibinfo  {journal} {Phys. Rev. D}\ }\textbf {\bibinfo {volume} {78}},\ \bibinfo {pages} {083507} (\bibinfo {year} {2008})},\ \Eprint {https://arxiv.org/abs/0807.1726} {arXiv:0807.1726 [astro-ph]} \BibitemShut {NoStop}%
\bibitem [{\citenamefont {Komatsu}\ \emph {et~al.}(2009)\citenamefont {Komatsu} \emph {et~al.}}]{WMAP:2008lyn}%
  \BibitemOpen
  \bibfield  {author} {\bibinfo {author} {\bibfnamefont {E.}~\bibnamefont {Komatsu}} \emph {et~al.} (\bibinfo {collaboration} {WMAP}),\ }\href {https://doi.org/10.1088/0067-0049/180/2/330} {\bibfield  {journal} {\bibinfo  {journal} {Astrophys. J. Suppl.}\ }\textbf {\bibinfo {volume} {180}},\ \bibinfo {pages} {330} (\bibinfo {year} {2009})},\ \Eprint {https://arxiv.org/abs/0803.0547} {arXiv:0803.0547 [astro-ph]} \BibitemShut {NoStop}%
\bibitem [{\citenamefont {Marsh}\ \emph {et~al.}(2013)\citenamefont {Marsh}, \citenamefont {Grin}, \citenamefont {Hlozek},\ and\ \citenamefont {Ferreira}}]{Marsh:2013taa}%
  \BibitemOpen
  \bibfield  {author} {\bibinfo {author} {\bibfnamefont {D.~J.~E.}\ \bibnamefont {Marsh}}, \bibinfo {author} {\bibfnamefont {D.}~\bibnamefont {Grin}}, \bibinfo {author} {\bibfnamefont {R.}~\bibnamefont {Hlozek}},\ and\ \bibinfo {author} {\bibfnamefont {P.~G.}\ \bibnamefont {Ferreira}},\ }\href {https://doi.org/10.1103/PhysRevD.87.121701} {\bibfield  {journal} {\bibinfo  {journal} {Phys. Rev. D}\ }\textbf {\bibinfo {volume} {87}},\ \bibinfo {pages} {121701} (\bibinfo {year} {2013})},\ \Eprint {https://arxiv.org/abs/1303.3008} {arXiv:1303.3008 [astro-ph.CO]} \BibitemShut {NoStop}%
\bibitem [{\citenamefont {Marsh}\ \emph {et~al.}(2014)\citenamefont {Marsh}, \citenamefont {Grin}, \citenamefont {Hlozek},\ and\ \citenamefont {Ferreira}}]{Marsh:2014qoa}%
  \BibitemOpen
  \bibfield  {author} {\bibinfo {author} {\bibfnamefont {D.~J.~E.}\ \bibnamefont {Marsh}}, \bibinfo {author} {\bibfnamefont {D.}~\bibnamefont {Grin}}, \bibinfo {author} {\bibfnamefont {R.}~\bibnamefont {Hlozek}},\ and\ \bibinfo {author} {\bibfnamefont {P.~G.}\ \bibnamefont {Ferreira}},\ }\href {https://doi.org/10.1103/PhysRevLett.113.011801} {\bibfield  {journal} {\bibinfo  {journal} {Phys. Rev. Lett.}\ }\textbf {\bibinfo {volume} {113}},\ \bibinfo {pages} {011801} (\bibinfo {year} {2014})},\ \Eprint {https://arxiv.org/abs/1403.4216} {arXiv:1403.4216 [astro-ph.CO]} \BibitemShut {NoStop}%
\bibitem [{\citenamefont {Visinelli}\ and\ \citenamefont {Gondolo}(2014)}]{Visinelli:2014twa}%
  \BibitemOpen
  \bibfield  {author} {\bibinfo {author} {\bibfnamefont {L.}~\bibnamefont {Visinelli}}\ and\ \bibinfo {author} {\bibfnamefont {P.}~\bibnamefont {Gondolo}},\ }\href {https://doi.org/10.1103/PhysRevLett.113.011802} {\bibfield  {journal} {\bibinfo  {journal} {Phys. Rev. Lett.}\ }\textbf {\bibinfo {volume} {113}},\ \bibinfo {pages} {011802} (\bibinfo {year} {2014})},\ \Eprint {https://arxiv.org/abs/1403.4594} {arXiv:1403.4594 [hep-ph]} \BibitemShut {NoStop}%
\bibitem [{\citenamefont {Hlozek}\ \emph {et~al.}(2018)\citenamefont {Hlozek}, \citenamefont {Marsh},\ and\ \citenamefont {Grin}}]{Hlozek:2017zzf}%
  \BibitemOpen
  \bibfield  {author} {\bibinfo {author} {\bibfnamefont {R.}~\bibnamefont {Hlozek}}, \bibinfo {author} {\bibfnamefont {D.~J.~E.}\ \bibnamefont {Marsh}},\ and\ \bibinfo {author} {\bibfnamefont {D.}~\bibnamefont {Grin}},\ }\href {https://doi.org/10.1093/mnras/sty271} {\bibfield  {journal} {\bibinfo  {journal} {Mon. Not. Roy. Astron. Soc.}\ }\textbf {\bibinfo {volume} {476}},\ \bibinfo {pages} {3063} (\bibinfo {year} {2018})},\ \Eprint {https://arxiv.org/abs/1708.05681} {arXiv:1708.05681 [astro-ph.CO]} \BibitemShut {NoStop}%
\bibitem [{\citenamefont {{Boylan-Kolchin}}\ \emph {et~al.}(2011)\citenamefont {{Boylan-Kolchin}}, \citenamefont {{Bullock}},\ and\ \citenamefont {{Kaplinghat}}}]{2011MNRAS.415L..40B}%
  \BibitemOpen
  \bibfield  {author} {\bibinfo {author} {\bibfnamefont {M.}~\bibnamefont {{Boylan-Kolchin}}}, \bibinfo {author} {\bibfnamefont {J.~S.}\ \bibnamefont {{Bullock}}},\ and\ \bibinfo {author} {\bibfnamefont {M.}~\bibnamefont {{Kaplinghat}}},\ }\href {https://doi.org/10.1111/j.1745-3933.2011.01074.x} {\bibfield  {journal} {\bibinfo  {journal} {MNRAS}\ }\textbf {\bibinfo {volume} {415}},\ \bibinfo {pages} {L40} (\bibinfo {year} {2011})},\ \Eprint {https://arxiv.org/abs/1103.0007} {arXiv:1103.0007 [astro-ph.CO]} \BibitemShut {NoStop}%
\bibitem [{\citenamefont {Hlozek}\ \emph {et~al.}(2015)\citenamefont {Hlozek}, \citenamefont {Grin}, \citenamefont {Marsh},\ and\ \citenamefont {Ferreira}}]{Hlozek:2014lca}%
  \BibitemOpen
  \bibfield  {author} {\bibinfo {author} {\bibfnamefont {R.}~\bibnamefont {Hlozek}}, \bibinfo {author} {\bibfnamefont {D.}~\bibnamefont {Grin}}, \bibinfo {author} {\bibfnamefont {D.~J.~E.}\ \bibnamefont {Marsh}},\ and\ \bibinfo {author} {\bibfnamefont {P.~G.}\ \bibnamefont {Ferreira}},\ }\href {https://doi.org/10.1103/PhysRevD.91.103512} {\bibfield  {journal} {\bibinfo  {journal} {Phys. Rev. D}\ }\textbf {\bibinfo {volume} {91}},\ \bibinfo {pages} {103512} (\bibinfo {year} {2015})},\ \Eprint {https://arxiv.org/abs/1410.2896} {arXiv:1410.2896 [astro-ph.CO]} \BibitemShut {NoStop}%
\bibitem [{\citenamefont {{Papastergis}}\ and\ \citenamefont {{Shankar}}(2016)}]{2016A&A...591A..58P}%
  \BibitemOpen
  \bibfield  {author} {\bibinfo {author} {\bibfnamefont {E.}~\bibnamefont {{Papastergis}}}\ and\ \bibinfo {author} {\bibfnamefont {F.}~\bibnamefont {{Shankar}}},\ }\href {https://doi.org/10.1051/0004-6361/201527854} {\bibfield  {journal} {\bibinfo  {journal} {Astron. Astrophys.}\ }\textbf {\bibinfo {volume} {591}},\ \bibinfo {eid} {A58} (\bibinfo {year} {2016})},\ \Eprint {https://arxiv.org/abs/1511.08741} {arXiv:1511.08741 [astro-ph.GA]} \BibitemShut {NoStop}%
\bibitem [{\citenamefont {Hui}\ \emph {et~al.}(2017)\citenamefont {Hui}, \citenamefont {Ostriker}, \citenamefont {Tremaine},\ and\ \citenamefont {Witten}}]{Hui:2016ltb}%
  \BibitemOpen
  \bibfield  {author} {\bibinfo {author} {\bibfnamefont {L.}~\bibnamefont {Hui}}, \bibinfo {author} {\bibfnamefont {J.~P.}\ \bibnamefont {Ostriker}}, \bibinfo {author} {\bibfnamefont {S.}~\bibnamefont {Tremaine}},\ and\ \bibinfo {author} {\bibfnamefont {E.}~\bibnamefont {Witten}},\ }\href {https://doi.org/10.1103/PhysRevD.95.043541} {\bibfield  {journal} {\bibinfo  {journal} {Phys. Rev. D}\ }\textbf {\bibinfo {volume} {95}},\ \bibinfo {pages} {043541} (\bibinfo {year} {2017})},\ \Eprint {https://arxiv.org/abs/1610.08297} {arXiv:1610.08297 [astro-ph.CO]} \BibitemShut {NoStop}%
\bibitem [{\citenamefont {Ir{\v{s}}i{\v{c}}}\ \emph {et~al.}(2017)\citenamefont {Ir{\v{s}}i{\v{c}}}, \citenamefont {Viel}, \citenamefont {Haehnelt}, \citenamefont {Bolton},\ and\ \citenamefont {Becker}}]{Irsic:2017yje}%
  \BibitemOpen
  \bibfield  {author} {\bibinfo {author} {\bibfnamefont {V.}~\bibnamefont {Ir{\v{s}}i{\v{c}}}}, \bibinfo {author} {\bibfnamefont {M.}~\bibnamefont {Viel}}, \bibinfo {author} {\bibfnamefont {M.~G.}\ \bibnamefont {Haehnelt}}, \bibinfo {author} {\bibfnamefont {J.~S.}\ \bibnamefont {Bolton}},\ and\ \bibinfo {author} {\bibfnamefont {G.~D.}\ \bibnamefont {Becker}},\ }\href {https://doi.org/10.1103/PhysRevLett.119.031302} {\bibfield  {journal} {\bibinfo  {journal} {Phys. Rev. Lett.}\ }\textbf {\bibinfo {volume} {119}},\ \bibinfo {pages} {031302} (\bibinfo {year} {2017})},\ \Eprint {https://arxiv.org/abs/1703.04683} {arXiv:1703.04683 [astro-ph.CO]} \BibitemShut {NoStop}%
\bibitem [{\citenamefont {{Bar-Or}}\ \emph {et~al.}(2019)\citenamefont {{Bar-Or}}, \citenamefont {{Fouvry}},\ and\ \citenamefont {{Tremaine}}}]{2019ApJ...871...28B}%
  \BibitemOpen
  \bibfield  {author} {\bibinfo {author} {\bibfnamefont {B.}~\bibnamefont {{Bar-Or}}}, \bibinfo {author} {\bibfnamefont {J.-B.}\ \bibnamefont {{Fouvry}}},\ and\ \bibinfo {author} {\bibfnamefont {S.}~\bibnamefont {{Tremaine}}},\ }\href {https://doi.org/10.3847/1538-4357/aaf28c} {\bibfield  {journal} {\bibinfo  {journal} {\apj}\ }\textbf {\bibinfo {volume} {871}},\ \bibinfo {eid} {28} (\bibinfo {year} {2019})},\ \Eprint {https://arxiv.org/abs/1809.07673} {arXiv:1809.07673 [astro-ph.GA]} \BibitemShut {NoStop}%
\bibitem [{\citenamefont {Rogers}\ and\ \citenamefont {Peiris}(2021)}]{Rogers:2020ltq}%
  \BibitemOpen
  \bibfield  {author} {\bibinfo {author} {\bibfnamefont {K.~K.}\ \bibnamefont {Rogers}}\ and\ \bibinfo {author} {\bibfnamefont {H.~V.}\ \bibnamefont {Peiris}},\ }\href {https://doi.org/10.1103/PhysRevLett.126.071302} {\bibfield  {journal} {\bibinfo  {journal} {Phys. Rev. Lett.}\ }\textbf {\bibinfo {volume} {126}},\ \bibinfo {pages} {071302} (\bibinfo {year} {2021})},\ \Eprint {https://arxiv.org/abs/2007.12705} {arXiv:2007.12705 [astro-ph.CO]} \BibitemShut {NoStop}%
\bibitem [{\citenamefont {Dalal}\ \emph {et~al.}(2021)\citenamefont {Dalal}, \citenamefont {Bovy}, \citenamefont {Hui},\ and\ \citenamefont {Li}}]{Dalal:2020mjw}%
  \BibitemOpen
  \bibfield  {author} {\bibinfo {author} {\bibfnamefont {N.}~\bibnamefont {Dalal}}, \bibinfo {author} {\bibfnamefont {J.}~\bibnamefont {Bovy}}, \bibinfo {author} {\bibfnamefont {L.}~\bibnamefont {Hui}},\ and\ \bibinfo {author} {\bibfnamefont {X.}~\bibnamefont {Li}},\ }\href {https://doi.org/10.1088/1475-7516/2021/03/076} {\bibfield  {journal} {\bibinfo  {journal} {JCAP}\ }\textbf {\bibinfo {volume} {03}},\ \bibinfo {pages} {076}},\ \Eprint {https://arxiv.org/abs/2011.13141} {arXiv:2011.13141 [astro-ph.CO]} \BibitemShut {NoStop}%
\bibitem [{\citenamefont {Dentler}\ \emph {et~al.}(2022)\citenamefont {Dentler}, \citenamefont {Marsh}, \citenamefont {Hlo{\v{z}}ek}, \citenamefont {Lagu{\"e}}, \citenamefont {Rogers},\ and\ \citenamefont {Grin}}]{Dentler:2021zij}%
  \BibitemOpen
  \bibfield  {author} {\bibinfo {author} {\bibfnamefont {M.}~\bibnamefont {Dentler}}, \bibinfo {author} {\bibfnamefont {D.~J.~E.}\ \bibnamefont {Marsh}}, \bibinfo {author} {\bibfnamefont {R.}~\bibnamefont {Hlo{\v{z}}ek}}, \bibinfo {author} {\bibfnamefont {A.}~\bibnamefont {Lagu{\"e}}}, \bibinfo {author} {\bibfnamefont {K.~K.}\ \bibnamefont {Rogers}},\ and\ \bibinfo {author} {\bibfnamefont {D.}~\bibnamefont {Grin}},\ }\href {https://doi.org/10.1093/mnras/stac1946} {\bibfield  {journal} {\bibinfo  {journal} {Mon. Not. Roy. Astron. Soc.}\ }\textbf {\bibinfo {volume} {515}},\ \bibinfo {pages} {5646} (\bibinfo {year} {2022})},\ \Eprint {https://arxiv.org/abs/2111.01199} {arXiv:2111.01199 [astro-ph.CO]} \BibitemShut {NoStop}%
\bibitem [{\citenamefont {Lagu{\"e}}\ \emph {et~al.}(2022)\citenamefont {Lagu{\"e}}, \citenamefont {Bond}, \citenamefont {Hlo{\v{z}}ek}, \citenamefont {Rogers}, \citenamefont {Marsh},\ and\ \citenamefont {Grin}}]{Lague:2021frh}%
  \BibitemOpen
  \bibfield  {author} {\bibinfo {author} {\bibfnamefont {A.}~\bibnamefont {Lagu{\"e}}}, \bibinfo {author} {\bibfnamefont {J.~R.}\ \bibnamefont {Bond}}, \bibinfo {author} {\bibfnamefont {R.}~\bibnamefont {Hlo{\v{z}}ek}}, \bibinfo {author} {\bibfnamefont {K.~K.}\ \bibnamefont {Rogers}}, \bibinfo {author} {\bibfnamefont {D.~J.~E.}\ \bibnamefont {Marsh}},\ and\ \bibinfo {author} {\bibfnamefont {D.}~\bibnamefont {Grin}},\ }\href {https://doi.org/10.1088/1475-7516/2022/01/049} {\bibfield  {journal} {\bibinfo  {journal} {JCAP}\ }\textbf {\bibinfo {volume} {01}}\bibfield  {number} {\bibinfo  {number} { (01)},\ \bibinfo {pages} {049}},\ }\Eprint {https://arxiv.org/abs/2104.07802} {arXiv:2104.07802 [astro-ph.CO]} \BibitemShut {NoStop}%
\bibitem [{\citenamefont {Rogers}\ \emph {et~al.}(2023)\citenamefont {Rogers}, \citenamefont {Hlo{\v{z}}ek}, \citenamefont {Lagu{\"e}}, \citenamefont {Ivanov}, \citenamefont {Philcox}, \citenamefont {Cabass}, \citenamefont {Akitsu},\ and\ \citenamefont {Marsh}}]{Rogers:2023ezo}%
  \BibitemOpen
  \bibfield  {author} {\bibinfo {author} {\bibfnamefont {K.~K.}\ \bibnamefont {Rogers}}, \bibinfo {author} {\bibfnamefont {R.}~\bibnamefont {Hlo{\v{z}}ek}}, \bibinfo {author} {\bibfnamefont {A.}~\bibnamefont {Lagu{\"e}}}, \bibinfo {author} {\bibfnamefont {M.~M.}\ \bibnamefont {Ivanov}}, \bibinfo {author} {\bibfnamefont {O.~H.~E.}\ \bibnamefont {Philcox}}, \bibinfo {author} {\bibfnamefont {G.}~\bibnamefont {Cabass}}, \bibinfo {author} {\bibfnamefont {K.}~\bibnamefont {Akitsu}},\ and\ \bibinfo {author} {\bibfnamefont {D.~J.~E.}\ \bibnamefont {Marsh}},\ }\href {https://doi.org/10.1088/1475-7516/2023/06/023} {\bibfield  {journal} {\bibinfo  {journal} {JCAP}\ }\textbf {\bibinfo {volume} {06}},\ \bibinfo {pages} {023}},\ \Eprint {https://arxiv.org/abs/2301.08361} {arXiv:2301.08361 [astro-ph.CO]} \BibitemShut {NoStop}%
\bibitem [{\citenamefont {Winch}\ \emph {et~al.}(2024)\citenamefont {Winch}, \citenamefont {Rogers}, \citenamefont {Hlo{\v{z}}ek},\ and\ \citenamefont {Marsh}}]{Winch:2024mrt}%
  \BibitemOpen
  \bibfield  {author} {\bibinfo {author} {\bibfnamefont {H.}~\bibnamefont {Winch}}, \bibinfo {author} {\bibfnamefont {K.~K.}\ \bibnamefont {Rogers}}, \bibinfo {author} {\bibfnamefont {R.}~\bibnamefont {Hlo{\v{z}}ek}},\ and\ \bibinfo {author} {\bibfnamefont {D.~J.~E.}\ \bibnamefont {Marsh}},\ }\href {https://doi.org/10.3847/1538-4357/ad7a73} {\bibfield  {journal} {\bibinfo  {journal} {Astrophys. J.}\ }\textbf {\bibinfo {volume} {976}},\ \bibinfo {pages} {40} (\bibinfo {year} {2024})},\ \Eprint {https://arxiv.org/abs/2404.11071} {arXiv:2404.11071 [astro-ph.CO]} \BibitemShut {NoStop}%
\bibitem [{\citenamefont {Sipple}\ \emph {et~al.}(2025)\citenamefont {Sipple}, \citenamefont {Lidz}, \citenamefont {Grin},\ and\ \citenamefont {Sun}}]{Sipple:2024svt}%
  \BibitemOpen
  \bibfield  {author} {\bibinfo {author} {\bibfnamefont {J.}~\bibnamefont {Sipple}}, \bibinfo {author} {\bibfnamefont {A.}~\bibnamefont {Lidz}}, \bibinfo {author} {\bibfnamefont {D.}~\bibnamefont {Grin}},\ and\ \bibinfo {author} {\bibfnamefont {G.}~\bibnamefont {Sun}},\ }\href {https://doi.org/10.1093/mnras/staf340} {\bibfield  {journal} {\bibinfo  {journal} {Mon. Not. Roy. Astron. Soc.}\ }\textbf {\bibinfo {volume} {538}},\ \bibinfo {pages} {1830} (\bibinfo {year} {2025})},\ \Eprint {https://arxiv.org/abs/2407.17059} {arXiv:2407.17059 [astro-ph.CO]} \BibitemShut {NoStop}%
\bibitem [{\citenamefont {Akrami}\ \emph {et~al.}(2020)\citenamefont {Akrami} \emph {et~al.}}]{Planck:2018jri}%
  \BibitemOpen
  \bibfield  {author} {\bibinfo {author} {\bibfnamefont {Y.}~\bibnamefont {Akrami}} \emph {et~al.} (\bibinfo {collaboration} {Planck}),\ }\href {https://doi.org/10.1051/0004-6361/201833887} {\bibfield  {journal} {\bibinfo  {journal} {Astron. Astrophys.}\ }\textbf {\bibinfo {volume} {641}},\ \bibinfo {pages} {A10} (\bibinfo {year} {2020})},\ \Eprint {https://arxiv.org/abs/1807.06211} {arXiv:1807.06211 [astro-ph.CO]} \BibitemShut {NoStop}%
\bibitem [{\citenamefont {Gordon}\ and\ \citenamefont {Hu}(2004)}]{Gordon:2004ez}%
  \BibitemOpen
  \bibfield  {author} {\bibinfo {author} {\bibfnamefont {C.}~\bibnamefont {Gordon}}\ and\ \bibinfo {author} {\bibfnamefont {W.}~\bibnamefont {Hu}},\ }\href {https://doi.org/10.1103/PhysRevD.70.083003} {\bibfield  {journal} {\bibinfo  {journal} {Phys. Rev. D}\ }\textbf {\bibinfo {volume} {70}},\ \bibinfo {pages} {083003} (\bibinfo {year} {2004})},\ \Eprint {https://arxiv.org/abs/astro-ph/0406496} {arXiv:astro-ph/0406496} \BibitemShut {NoStop}%
\bibitem [{\citenamefont {Coble}\ \emph {et~al.}(1997)\citenamefont {Coble}, \citenamefont {Dodelson},\ and\ \citenamefont {Frieman}}]{Coble:1996te}%
  \BibitemOpen
  \bibfield  {author} {\bibinfo {author} {\bibfnamefont {K.}~\bibnamefont {Coble}}, \bibinfo {author} {\bibfnamefont {S.}~\bibnamefont {Dodelson}},\ and\ \bibinfo {author} {\bibfnamefont {J.~A.}\ \bibnamefont {Frieman}},\ }\href {https://doi.org/10.1103/PhysRevD.55.1851} {\bibfield  {journal} {\bibinfo  {journal} {Phys. Rev. D}\ }\textbf {\bibinfo {volume} {55}},\ \bibinfo {pages} {1851} (\bibinfo {year} {1997})},\ \Eprint {https://arxiv.org/abs/astro-ph/9608122} {arXiv:astro-ph/9608122} \BibitemShut {NoStop}%
\bibitem [{\citenamefont {Frieman}\ \emph {et~al.}(1995)\citenamefont {Frieman}, \citenamefont {Hill}, \citenamefont {Stebbins},\ and\ \citenamefont {Waga}}]{Frieman:1995pm}%
  \BibitemOpen
  \bibfield  {author} {\bibinfo {author} {\bibfnamefont {J.~A.}\ \bibnamefont {Frieman}}, \bibinfo {author} {\bibfnamefont {C.~T.}\ \bibnamefont {Hill}}, \bibinfo {author} {\bibfnamefont {A.}~\bibnamefont {Stebbins}},\ and\ \bibinfo {author} {\bibfnamefont {I.}~\bibnamefont {Waga}},\ }\href {https://doi.org/10.1103/PhysRevLett.75.2077} {\bibfield  {journal} {\bibinfo  {journal} {Phys. Rev. Lett.}\ }\textbf {\bibinfo {volume} {75}},\ \bibinfo {pages} {2077} (\bibinfo {year} {1995})},\ \Eprint {https://arxiv.org/abs/astro-ph/9505060} {arXiv:astro-ph/9505060} \BibitemShut {NoStop}%
\bibitem [{\citenamefont {Batell}\ \emph {et~al.}(2026)\citenamefont {Batell}, \citenamefont {Dasgupta}, \citenamefont {Dutta},\ and\ \citenamefont {Ghalsasi}}]{Batell:2026avi}%
  \BibitemOpen
  \bibfield  {author} {\bibinfo {author} {\bibfnamefont {B.}~\bibnamefont {Batell}}, \bibinfo {author} {\bibfnamefont {A.}~\bibnamefont {Dasgupta}}, \bibinfo {author} {\bibfnamefont {S.}~\bibnamefont {Dutta}},\ and\ \bibinfo {author} {\bibfnamefont {A.}~\bibnamefont {Ghalsasi}},\ }\href@noop {} {\  (\bibinfo {year} {2026})},\ \Eprint {https://arxiv.org/abs/2606.02706} {arXiv:2606.02706 [hep-ph]} \BibitemShut {NoStop}%
\bibitem [{\citenamefont {Cookmeyer}\ \emph {et~al.}(2020)\citenamefont {Cookmeyer}, \citenamefont {Grin},\ and\ \citenamefont {Smith}}]{Cookmeyer:2019rna}%
  \BibitemOpen
  \bibfield  {author} {\bibinfo {author} {\bibfnamefont {T.}~\bibnamefont {Cookmeyer}}, \bibinfo {author} {\bibfnamefont {D.}~\bibnamefont {Grin}},\ and\ \bibinfo {author} {\bibfnamefont {T.~L.}\ \bibnamefont {Smith}},\ }\href {https://doi.org/10.1103/PhysRevD.101.023501} {\bibfield  {journal} {\bibinfo  {journal} {Phys. Rev. D}\ }\textbf {\bibinfo {volume} {101}},\ \bibinfo {pages} {023501} (\bibinfo {year} {2020})},\ \Eprint {https://arxiv.org/abs/1909.11094} {arXiv:1909.11094 [astro-ph.CO]} \BibitemShut {NoStop}%
\bibitem [{\citenamefont {Passaglia}\ and\ \citenamefont {Hu}(2022)}]{Passaglia:2022bcr}%
  \BibitemOpen
  \bibfield  {author} {\bibinfo {author} {\bibfnamefont {S.}~\bibnamefont {Passaglia}}\ and\ \bibinfo {author} {\bibfnamefont {W.}~\bibnamefont {Hu}},\ }\href {https://doi.org/10.1103/PhysRevD.105.123529} {\bibfield  {journal} {\bibinfo  {journal} {Phys. Rev. D}\ }\textbf {\bibinfo {volume} {105}},\ \bibinfo {pages} {123529} (\bibinfo {year} {2022})},\ \Eprint {https://arxiv.org/abs/2201.10238} {arXiv:2201.10238 [astro-ph.CO]} \BibitemShut {NoStop}%
\bibitem [{\citenamefont {Ure{\~n}a-L{\'o}pez}\ and\ \citenamefont {Gonzalez-Morales}(2016)}]{Urena-Lopez:2015gur}%
  \BibitemOpen
  \bibfield  {author} {\bibinfo {author} {\bibfnamefont {L.~A.}\ \bibnamefont {Ure{\~n}a-L{\'o}pez}}\ and\ \bibinfo {author} {\bibfnamefont {A.~X.}\ \bibnamefont {Gonzalez-Morales}},\ }\href {https://doi.org/10.1088/1475-7516/2016/07/048} {\bibfield  {journal} {\bibinfo  {journal} {JCAP}\ }\textbf {\bibinfo {volume} {07}},\ \bibinfo {pages} {048}},\ \Eprint {https://arxiv.org/abs/1511.08195} {arXiv:1511.08195 [astro-ph.CO]} \BibitemShut {NoStop}%
\bibitem [{\citenamefont {Liu}\ \emph {et~al.}(2025{\natexlab{a}})\citenamefont {Liu}, \citenamefont {Hu},\ and\ \citenamefont {Grin}}]{Liu:2024yne}%
  \BibitemOpen
  \bibfield  {author} {\bibinfo {author} {\bibfnamefont {R.}~\bibnamefont {Liu}}, \bibinfo {author} {\bibfnamefont {W.}~\bibnamefont {Hu}},\ and\ \bibinfo {author} {\bibfnamefont {D.}~\bibnamefont {Grin}},\ }\href {https://doi.org/10.1103/1z4c-1w7f} {\bibfield  {journal} {\bibinfo  {journal} {Phys. Rev. D}\ }\textbf {\bibinfo {volume} {112}},\ \bibinfo {pages} {023513} (\bibinfo {year} {2025}{\natexlab{a}})},\ \Eprint {https://arxiv.org/abs/2412.15192} {arXiv:2412.15192 [astro-ph.CO]} \BibitemShut {NoStop}%
\bibitem [{\citenamefont {Baryakhtar}\ \emph {et~al.}(2024)\citenamefont {Baryakhtar}, \citenamefont {Simon},\ and\ \citenamefont {Weiner}}]{Baryakhtar:2024rky}%
  \BibitemOpen
  \bibfield  {author} {\bibinfo {author} {\bibfnamefont {M.}~\bibnamefont {Baryakhtar}}, \bibinfo {author} {\bibfnamefont {O.}~\bibnamefont {Simon}},\ and\ \bibinfo {author} {\bibfnamefont {Z.~J.}\ \bibnamefont {Weiner}},\ }\href {https://doi.org/10.1103/PhysRevD.110.083505} {\bibfield  {journal} {\bibinfo  {journal} {Phys. Rev. D}\ }\textbf {\bibinfo {volume} {110}},\ \bibinfo {pages} {083505} (\bibinfo {year} {2024})},\ \Eprint {https://arxiv.org/abs/2405.10358} {arXiv:2405.10358 [astro-ph.CO]} \BibitemShut {NoStop}%
\bibitem [{\citenamefont {Moss}\ \emph {et~al.}(2025)\citenamefont {Moss}, \citenamefont {Gaughan},\ and\ \citenamefont {Green}}]{Moss:2025ymr}%
  \BibitemOpen
  \bibfield  {author} {\bibinfo {author} {\bibfnamefont {A.}~\bibnamefont {Moss}}, \bibinfo {author} {\bibfnamefont {L.}~\bibnamefont {Gaughan}},\ and\ \bibinfo {author} {\bibfnamefont {A.~M.}\ \bibnamefont {Green}},\ }\href {https://doi.org/10.1103/fy1y-rjst} {\bibfield  {journal} {\bibinfo  {journal} {Phys. Rev. D}\ }\textbf {\bibinfo {volume} {111}},\ \bibinfo {pages} {123530} (\bibinfo {year} {2025})},\ \Eprint {https://arxiv.org/abs/2501.13662} {arXiv:2501.13662 [astro-ph.CO]} \BibitemShut {NoStop}%
\bibitem [{\citenamefont {{Gaughan}}\ \emph {et~al.}(2026)\citenamefont {{Gaughan}}, \citenamefont {{Green}},\ and\ \citenamefont {{Moss}}}]{2026arXiv260512054G}%
  \BibitemOpen
  \bibfield  {author} {\bibinfo {author} {\bibfnamefont {L.}~\bibnamefont {{Gaughan}}}, \bibinfo {author} {\bibfnamefont {A.~M.}\ \bibnamefont {{Green}}},\ and\ \bibinfo {author} {\bibfnamefont {A.}~\bibnamefont {{Moss}}},\ }\href@noop {} {\  (\bibinfo {year} {2026})},\ \Eprint {https://arxiv.org/abs/2605.12054} {arXiv:2605.12054 [astro-ph.CO]} \BibitemShut {NoStop}%
\bibitem [{\citenamefont {Liu}\ \emph {et~al.}(2026)\citenamefont {Liu}, \citenamefont {Zhu}, \citenamefont {Hu},\ and\ \citenamefont {Miranda}}]{Liu:2025bss}%
  \BibitemOpen
  \bibfield  {author} {\bibinfo {author} {\bibfnamefont {R.}~\bibnamefont {Liu}}, \bibinfo {author} {\bibfnamefont {Y.}~\bibnamefont {Zhu}}, \bibinfo {author} {\bibfnamefont {W.}~\bibnamefont {Hu}},\ and\ \bibinfo {author} {\bibfnamefont {V.}~\bibnamefont {Miranda}},\ }\href {https://doi.org/10.1103/3s1m-9zpc} {\bibfield  {journal} {\bibinfo  {journal} {Phys. Rev. D}\ }\textbf {\bibinfo {volume} {113}},\ \bibinfo {pages} {083506} (\bibinfo {year} {2026})},\ \bibinfo {note} {\href{https://github.com/SBU-COSMOLIKE/cocoa_axions}{code link}},\ \Eprint {https://arxiv.org/abs/2510.14957} {arXiv:2510.14957 [astro-ph.CO]} \BibitemShut {NoStop}%
\bibitem [{\citenamefont {{Lagu{\"e}}}\ \emph {et~al.}(2026)\citenamefont {{Lagu{\"e}}} \emph {et~al.}}]{2026arXiv260606410L}%
  \BibitemOpen
  \bibfield  {author} {\bibinfo {author} {\bibfnamefont {A.}~\bibnamefont {{Lagu{\"e}}}} \emph {et~al.},\ }\href@noop {} {\  (\bibinfo {year} {2026})},\ \Eprint {https://arxiv.org/abs/2606.06410} {arXiv:2606.06410 [astro-ph.CO]} \BibitemShut {NoStop}%
\bibitem [{\citenamefont {Aghanim}\ \emph {et~al.}(2020)\citenamefont {Aghanim} \emph {et~al.}}]{Aghanim:2018eyx}%
  \BibitemOpen
  \bibfield  {author} {\bibinfo {author} {\bibfnamefont {N.}~\bibnamefont {Aghanim}} \emph {et~al.} (\bibinfo {collaboration} {Planck}),\ }\href {https://doi.org/10.1051/0004-6361/201833910} {\bibfield  {journal} {\bibinfo  {journal} {Astron. Astrophys.}\ }\textbf {\bibinfo {volume} {641}},\ \bibinfo {pages} {A6} (\bibinfo {year} {2020})},\ \Eprint {https://arxiv.org/abs/1807.06209} {arXiv:1807.06209 [astro-ph.CO]} \BibitemShut {NoStop}%
\bibitem [{\citenamefont {Hu}\ and\ \citenamefont {Sugiyama}(1995)}]{Hu:1994jd}%
  \BibitemOpen
  \bibfield  {author} {\bibinfo {author} {\bibfnamefont {W.}~\bibnamefont {Hu}}\ and\ \bibinfo {author} {\bibfnamefont {N.}~\bibnamefont {Sugiyama}},\ }\href {https://doi.org/10.1103/PhysRevD.51.2599} {\bibfield  {journal} {\bibinfo  {journal} {Phys. Rev. D}\ }\textbf {\bibinfo {volume} {51}},\ \bibinfo {pages} {2599} (\bibinfo {year} {1995})},\ \Eprint {https://arxiv.org/abs/astro-ph/9411008} {arXiv:astro-ph/9411008} \BibitemShut {NoStop}%
\bibitem [{\citenamefont {Petretti}\ \emph {et~al.}(2026)\citenamefont {Petretti}, \citenamefont {Singh}, \citenamefont {Braglia}, \citenamefont {Chen}, \citenamefont {Fan},\ and\ \citenamefont {Li}}]{Petretti:2026ayw}%
  \BibitemOpen
  \bibfield  {author} {\bibinfo {author} {\bibfnamefont {C.}~\bibnamefont {Petretti}}, \bibinfo {author} {\bibfnamefont {P.}~\bibnamefont {Singh}}, \bibinfo {author} {\bibfnamefont {M.}~\bibnamefont {Braglia}}, \bibinfo {author} {\bibfnamefont {X.}~\bibnamefont {Chen}}, \bibinfo {author} {\bibfnamefont {J.}~\bibnamefont {Fan}},\ and\ \bibinfo {author} {\bibfnamefont {L.}~\bibnamefont {Li}},\ }\href@noop {} {\  (\bibinfo {year} {2026})},\ \Eprint {https://arxiv.org/abs/2606.11312} {arXiv:2606.11312 [astro-ph.CO]} \BibitemShut {NoStop}%
\bibitem [{\citenamefont {Ade}\ \emph {et~al.}(2021)\citenamefont {Ade} \emph {et~al.}}]{BICEP:2021xfz}%
  \BibitemOpen
  \bibfield  {author} {\bibinfo {author} {\bibfnamefont {P.~A.~R.}\ \bibnamefont {Ade}} \emph {et~al.} (\bibinfo {collaboration} {BICEP, Keck}),\ }\href {https://doi.org/10.1103/PhysRevLett.127.151301} {\bibfield  {journal} {\bibinfo  {journal} {Phys. Rev. Lett.}\ }\textbf {\bibinfo {volume} {127}},\ \bibinfo {pages} {151301} (\bibinfo {year} {2021})},\ \Eprint {https://arxiv.org/abs/2110.00483} {arXiv:2110.00483 [astro-ph.CO]} \BibitemShut {NoStop}%
\bibitem [{\citenamefont {{DES Collaboration}}(2026)}]{2026arXiv260210065D}%
  \BibitemOpen
  \bibfield  {author} {\bibinfo {author} {\bibnamefont {{DES Collaboration}}},\ }\href@noop {} {\  (\bibinfo {year} {2026})},\ \Eprint {https://arxiv.org/abs/2602.10065} {arXiv:2602.10065 [astro-ph.CO]} \BibitemShut {NoStop}%
\bibitem [{\citenamefont {Hu}\ \emph {et~al.}(2016)\citenamefont {Hu}, \citenamefont {Chiang}, \citenamefont {Li},\ and\ \citenamefont {LoVerde}}]{Hu:2016ssz}%
  \BibitemOpen
  \bibfield  {author} {\bibinfo {author} {\bibfnamefont {W.}~\bibnamefont {Hu}}, \bibinfo {author} {\bibfnamefont {C.-T.}\ \bibnamefont {Chiang}}, \bibinfo {author} {\bibfnamefont {Y.}~\bibnamefont {Li}},\ and\ \bibinfo {author} {\bibfnamefont {M.}~\bibnamefont {LoVerde}},\ }\href {https://doi.org/10.1103/PhysRevD.94.023002} {\bibfield  {journal} {\bibinfo  {journal} {Phys. Rev. D}\ }\textbf {\bibinfo {volume} {94}},\ \bibinfo {pages} {023002} (\bibinfo {year} {2016})},\ \Eprint {https://arxiv.org/abs/1605.01412} {arXiv:1605.01412 [astro-ph.CO]} \BibitemShut {NoStop}%
\bibitem [{\citenamefont {Jeong}\ and\ \citenamefont {Takahashi}(2013)}]{Jeong:2013xta}%
  \BibitemOpen
  \bibfield  {author} {\bibinfo {author} {\bibfnamefont {K.~S.}\ \bibnamefont {Jeong}}\ and\ \bibinfo {author} {\bibfnamefont {F.}~\bibnamefont {Takahashi}},\ }\href {https://doi.org/10.1016/j.physletb.2013.10.061} {\bibfield  {journal} {\bibinfo  {journal} {Phys. Lett. B}\ }\textbf {\bibinfo {volume} {727}},\ \bibinfo {pages} {448} (\bibinfo {year} {2013})},\ \Eprint {https://arxiv.org/abs/1304.8131} {arXiv:1304.8131 [hep-ph]} \BibitemShut {NoStop}%
\bibitem [{\citenamefont {Nomura}\ \emph {et~al.}(2016)\citenamefont {Nomura}, \citenamefont {Rajendran},\ and\ \citenamefont {Sanches}}]{Nomura:2015xil}%
  \BibitemOpen
  \bibfield  {author} {\bibinfo {author} {\bibfnamefont {Y.}~\bibnamefont {Nomura}}, \bibinfo {author} {\bibfnamefont {S.}~\bibnamefont {Rajendran}},\ and\ \bibinfo {author} {\bibfnamefont {F.}~\bibnamefont {Sanches}},\ }\href {https://doi.org/10.1103/PhysRevLett.116.141803} {\bibfield  {journal} {\bibinfo  {journal} {Phys. Rev. Lett.}\ }\textbf {\bibinfo {volume} {116}},\ \bibinfo {pages} {141803} (\bibinfo {year} {2016})},\ \Eprint {https://arxiv.org/abs/1511.06347} {arXiv:1511.06347 [hep-ph]} \BibitemShut {NoStop}%
\bibitem [{\citenamefont {Kawasaki}\ \emph {et~al.}(2016)\citenamefont {Kawasaki}, \citenamefont {Takahashi},\ and\ \citenamefont {Yamada}}]{Kawasaki:2015lpf}%
  \BibitemOpen
  \bibfield  {author} {\bibinfo {author} {\bibfnamefont {M.}~\bibnamefont {Kawasaki}}, \bibinfo {author} {\bibfnamefont {F.}~\bibnamefont {Takahashi}},\ and\ \bibinfo {author} {\bibfnamefont {M.}~\bibnamefont {Yamada}},\ }\href {https://doi.org/10.1016/j.physletb.2015.12.075} {\bibfield  {journal} {\bibinfo  {journal} {Phys. Lett. B}\ }\textbf {\bibinfo {volume} {753}},\ \bibinfo {pages} {677} (\bibinfo {year} {2016})},\ \Eprint {https://arxiv.org/abs/1511.05030} {arXiv:1511.05030 [hep-ph]} \BibitemShut {NoStop}%
\bibitem [{\citenamefont {Takahashi}\ and\ \citenamefont {Yamada}(2015)}]{Takahashi:2015waa}%
  \BibitemOpen
  \bibfield  {author} {\bibinfo {author} {\bibfnamefont {F.}~\bibnamefont {Takahashi}}\ and\ \bibinfo {author} {\bibfnamefont {M.}~\bibnamefont {Yamada}},\ }\href {https://doi.org/10.1088/1475-7516/2015/10/010} {\bibfield  {journal} {\bibinfo  {journal} {JCAP}\ }\textbf {\bibinfo {volume} {10}},\ \bibinfo {pages} {010}},\ \Eprint {https://arxiv.org/abs/1507.06387} {arXiv:1507.06387 [hep-ph]} \BibitemShut {NoStop}%
\bibitem [{\citenamefont {Nakayama}\ and\ \citenamefont {Takimoto}(2015)}]{Nakayama:2015pba}%
  \BibitemOpen
  \bibfield  {author} {\bibinfo {author} {\bibfnamefont {K.}~\bibnamefont {Nakayama}}\ and\ \bibinfo {author} {\bibfnamefont {M.}~\bibnamefont {Takimoto}},\ }\href {https://doi.org/10.1016/j.physletb.2015.07.001} {\bibfield  {journal} {\bibinfo  {journal} {Phys. Lett. B}\ }\textbf {\bibinfo {volume} {748}},\ \bibinfo {pages} {108} (\bibinfo {year} {2015})},\ \Eprint {https://arxiv.org/abs/1505.02119} {arXiv:1505.02119 [hep-ph]} \BibitemShut {NoStop}%
\bibitem [{\citenamefont {Graham}\ and\ \citenamefont {Racco}(2025)}]{Graham:2025iwx}%
  \BibitemOpen
  \bibfield  {author} {\bibinfo {author} {\bibfnamefont {P.~W.}\ \bibnamefont {Graham}}\ and\ \bibinfo {author} {\bibfnamefont {D.}~\bibnamefont {Racco}},\ }\href {https://doi.org/10.1007/JHEP12(2025)028} {\bibfield  {journal} {\bibinfo  {journal} {JHEP}\ }\textbf {\bibinfo {volume} {12}},\ \bibinfo {pages} {028}},\ \Eprint {https://arxiv.org/abs/2506.03348} {arXiv:2506.03348 [hep-ph]} \BibitemShut {NoStop}%
\bibitem [{\citenamefont {Harigaya}\ \emph {et~al.}(2015)\citenamefont {Harigaya}, \citenamefont {Ibe}, \citenamefont {Kawasaki},\ and\ \citenamefont {Yanagida}}]{Harigaya:2015hha}%
  \BibitemOpen
  \bibfield  {author} {\bibinfo {author} {\bibfnamefont {K.}~\bibnamefont {Harigaya}}, \bibinfo {author} {\bibfnamefont {M.}~\bibnamefont {Ibe}}, \bibinfo {author} {\bibfnamefont {M.}~\bibnamefont {Kawasaki}},\ and\ \bibinfo {author} {\bibfnamefont {T.~T.}\ \bibnamefont {Yanagida}},\ }\href {https://doi.org/10.1088/1475-7516/2015/11/003} {\bibfield  {journal} {\bibinfo  {journal} {JCAP}\ }\textbf {\bibinfo {volume} {11}},\ \bibinfo {pages} {003}},\ \Eprint {https://arxiv.org/abs/1507.00119} {arXiv:1507.00119 [hep-ph]} \BibitemShut {NoStop}%
\bibitem [{\citenamefont {Kobayashi}\ \emph {et~al.}(2013)\citenamefont {Kobayashi}, \citenamefont {Kurematsu},\ and\ \citenamefont {Takahashi}}]{Kobayashi:2013nva}%
  \BibitemOpen
  \bibfield  {author} {\bibinfo {author} {\bibfnamefont {T.}~\bibnamefont {Kobayashi}}, \bibinfo {author} {\bibfnamefont {R.}~\bibnamefont {Kurematsu}},\ and\ \bibinfo {author} {\bibfnamefont {F.}~\bibnamefont {Takahashi}},\ }\href {https://doi.org/10.1088/1475-7516/2013/09/032} {\bibfield  {journal} {\bibinfo  {journal} {JCAP}\ }\textbf {\bibinfo {volume} {09}},\ \bibinfo {pages} {032}},\ \Eprint {https://arxiv.org/abs/1304.0922} {arXiv:1304.0922 [hep-ph]} \BibitemShut {NoStop}%
\bibitem [{\citenamefont {Chung}\ and\ \citenamefont {Tadepalli}(2024)}]{Chung:2023xcv}%
  \BibitemOpen
  \bibfield  {author} {\bibinfo {author} {\bibfnamefont {D.~J.~H.}\ \bibnamefont {Chung}}\ and\ \bibinfo {author} {\bibfnamefont {S.~C.}\ \bibnamefont {Tadepalli}},\ }\href {https://doi.org/10.1103/PhysRevD.109.023539} {\bibfield  {journal} {\bibinfo  {journal} {Phys. Rev. D}\ }\textbf {\bibinfo {volume} {109}},\ \bibinfo {pages} {023539} (\bibinfo {year} {2024})},\ \Eprint {https://arxiv.org/abs/2309.17010} {arXiv:2309.17010 [astro-ph.CO]} \BibitemShut {NoStop}%
\bibitem [{\citenamefont {Ballesteros}\ \emph {et~al.}(2021)\citenamefont {Ballesteros}, \citenamefont {Ringwald}, \citenamefont {Tamarit},\ and\ \citenamefont {Welling}}]{Ballesteros:2021bee}%
  \BibitemOpen
  \bibfield  {author} {\bibinfo {author} {\bibfnamefont {G.}~\bibnamefont {Ballesteros}}, \bibinfo {author} {\bibfnamefont {A.}~\bibnamefont {Ringwald}}, \bibinfo {author} {\bibfnamefont {C.}~\bibnamefont {Tamarit}},\ and\ \bibinfo {author} {\bibfnamefont {Y.}~\bibnamefont {Welling}},\ }\href {https://doi.org/10.1088/1475-7516/2021/09/036} {\bibfield  {journal} {\bibinfo  {journal} {JCAP}\ }\textbf {\bibinfo {volume} {09}},\ \bibinfo {pages} {036}},\ \Eprint {https://arxiv.org/abs/2104.13847} {arXiv:2104.13847 [hep-ph]} \BibitemShut {NoStop}%
\bibitem [{\citenamefont {Caputo}\ \emph {et~al.}(2024)\citenamefont {Caputo}, \citenamefont {Geller},\ and\ \citenamefont {Rossi}}]{Caputo:2023ikd}%
  \BibitemOpen
  \bibfield  {author} {\bibinfo {author} {\bibfnamefont {A.}~\bibnamefont {Caputo}}, \bibinfo {author} {\bibfnamefont {M.}~\bibnamefont {Geller}},\ and\ \bibinfo {author} {\bibfnamefont {G.}~\bibnamefont {Rossi}},\ }\href {https://doi.org/10.1103/PhysRevD.110.055027} {\bibfield  {journal} {\bibinfo  {journal} {Phys. Rev. D}\ }\textbf {\bibinfo {volume} {110}},\ \bibinfo {pages} {055027} (\bibinfo {year} {2024})},\ \Eprint {https://arxiv.org/abs/2306.00056} {arXiv:2306.00056 [hep-ph]} \BibitemShut {NoStop}%
\bibitem [{\citenamefont {Dal~Cin}\ and\ \citenamefont {Kobayashi}(2023)}]{DalCin:2023uai}%
  \BibitemOpen
  \bibfield  {author} {\bibinfo {author} {\bibfnamefont {D.}~\bibnamefont {Dal~Cin}}\ and\ \bibinfo {author} {\bibfnamefont {T.}~\bibnamefont {Kobayashi}},\ }\href {https://doi.org/10.1103/PhysRevD.108.063530} {\bibfield  {journal} {\bibinfo  {journal} {Phys. Rev. D}\ }\textbf {\bibinfo {volume} {108}},\ \bibinfo {pages} {063530} (\bibinfo {year} {2023})},\ \Eprint {https://arxiv.org/abs/2305.18524} {arXiv:2305.18524 [hep-ph]} \BibitemShut {NoStop}%
\bibitem [{\citenamefont {Takahashi}\ and\ \citenamefont {Yin}(2019)}]{Takahashi:2019pqf}%
  \BibitemOpen
  \bibfield  {author} {\bibinfo {author} {\bibfnamefont {F.}~\bibnamefont {Takahashi}}\ and\ \bibinfo {author} {\bibfnamefont {W.}~\bibnamefont {Yin}},\ }\href {https://doi.org/10.1007/JHEP10(2019)120} {\bibfield  {journal} {\bibinfo  {journal} {JHEP}\ }\textbf {\bibinfo {volume} {10}},\ \bibinfo {pages} {120}},\ \Eprint {https://arxiv.org/abs/1908.06071} {arXiv:1908.06071 [hep-ph]} \BibitemShut {NoStop}%
\bibitem [{\citenamefont {Takahashi}\ \emph {et~al.}(2012)\citenamefont {Takahashi}, \citenamefont {Sato}, \citenamefont {Nishimichi}, \citenamefont {Taruya},\ and\ \citenamefont {Oguri}}]{Takahashi:2012em}%
  \BibitemOpen
  \bibfield  {author} {\bibinfo {author} {\bibfnamefont {R.}~\bibnamefont {Takahashi}}, \bibinfo {author} {\bibfnamefont {M.}~\bibnamefont {Sato}}, \bibinfo {author} {\bibfnamefont {T.}~\bibnamefont {Nishimichi}}, \bibinfo {author} {\bibfnamefont {A.}~\bibnamefont {Taruya}},\ and\ \bibinfo {author} {\bibfnamefont {M.}~\bibnamefont {Oguri}},\ }\href {https://doi.org/10.1088/0004-637X/761/2/152} {\bibfield  {journal} {\bibinfo  {journal} {Astrophys. J.}\ }\textbf {\bibinfo {volume} {761}},\ \bibinfo {pages} {152} (\bibinfo {year} {2012})},\ \Eprint {https://arxiv.org/abs/1208.2701} {arXiv:1208.2701 [astro-ph.CO]} \BibitemShut {NoStop}%
\bibitem [{\citenamefont {Liu}\ \emph {et~al.}(2025{\natexlab{b}})\citenamefont {Liu}, \citenamefont {Hu},\ and\ \citenamefont {Xiao}}]{Liu:2025lts}%
  \BibitemOpen
  \bibfield  {author} {\bibinfo {author} {\bibfnamefont {R.}~\bibnamefont {Liu}}, \bibinfo {author} {\bibfnamefont {W.}~\bibnamefont {Hu}},\ and\ \bibinfo {author} {\bibfnamefont {H.}~\bibnamefont {Xiao}},\ }\href {https://doi.org/10.1103/7mcv-ltjg} {\bibfield  {journal} {\bibinfo  {journal} {Phys. Rev. D}\ }\textbf {\bibinfo {volume} {112}},\ \bibinfo {pages} {023552} (\bibinfo {year} {2025}{\natexlab{b}})},\ \Eprint {https://arxiv.org/abs/2504.01937} {arXiv:2504.01937 [astro-ph.CO]} \BibitemShut {NoStop}%
\bibitem [{\citenamefont {Amin}\ \emph {et~al.}(2025)\citenamefont {Amin}, \citenamefont {May},\ and\ \citenamefont {Mirbabayi}}]{Amin:2025sla}%
  \BibitemOpen
  \bibfield  {author} {\bibinfo {author} {\bibfnamefont {M.~A.}\ \bibnamefont {Amin}}, \bibinfo {author} {\bibfnamefont {S.}~\bibnamefont {May}},\ and\ \bibinfo {author} {\bibfnamefont {M.}~\bibnamefont {Mirbabayi}},\ }\href {https://doi.org/10.1088/1475-7516/2025/10/040} {\bibfield  {journal} {\bibinfo  {journal} {JCAP}\ }\textbf {\bibinfo {volume} {10}},\ \bibinfo {pages} {040}},\ \Eprint {https://arxiv.org/abs/2506.12131} {arXiv:2506.12131 [astro-ph.CO]} \BibitemShut {NoStop}%
\bibitem [{\citenamefont {Harigaya}\ \emph {et~al.}(2025)\citenamefont {Harigaya}, \citenamefont {Hu}, \citenamefont {Liu},\ and\ \citenamefont {Xiao}}]{Harigaya:2025pox}%
  \BibitemOpen
  \bibfield  {author} {\bibinfo {author} {\bibfnamefont {K.}~\bibnamefont {Harigaya}}, \bibinfo {author} {\bibfnamefont {W.}~\bibnamefont {Hu}}, \bibinfo {author} {\bibfnamefont {R.}~\bibnamefont {Liu}},\ and\ \bibinfo {author} {\bibfnamefont {H.}~\bibnamefont {Xiao}},\ }\href {https://doi.org/10.1103/xgf4-xqjh} {\bibfield  {journal} {\bibinfo  {journal} {Phys. Rev. D}\ }\textbf {\bibinfo {volume} {112}},\ \bibinfo {pages} {063554} (\bibinfo {year} {2025})},\ \Eprint {https://arxiv.org/abs/2507.01956} {arXiv:2507.01956 [astro-ph.CO]} \BibitemShut {NoStop}%
\end{thebibliography}%

\end{document}